\documentclass[journal=jpcbfk,manuscript=article]{achemso}

\usepackage{chemformula} 
\usepackage[T1]{fontenc} 

\usepackage{color}
\usepackage{amsthm}
\usepackage{graphicx}
\usepackage{threeparttable}
\usepackage{algorithm, algorithmicx}
\usepackage{algcompatible}

\newlength\myindent 
\author{Patrice Koehl}
\email{koehl@cs.ucdavis.edu}
\affiliation[Davis]
{Department of Computer Science and Genome Center, University of California, Davis, CA~~95616, USA}
\author{Marc Delarue}
\email{delarue@pasteur.fr}
\affiliation[Pasteur]
{Architecture et Dynamique des Macromol\'{e}cules Biologiques, Department of Structural Biology and Chemistry,
        UMR 3528 du CNRS, Institut Pasteur, 75015 Paris, France}
\author{Henri Orland}
\email{henri.orland@ipht.fr}
\affiliation[CEA]
{Institut de Physique Th\'{e}orique, Universit\'{e} Paris-Saclay,
               CEA, 91191 Gif/Yvette Cedex, France}

\title[AquaVit]
{Simultaneous identification of multiple binding sites in proteins:\\ A statistical mechanics approach}


\begin{document}


\begin{tocentry}
\includegraphics[height=3.45cm]{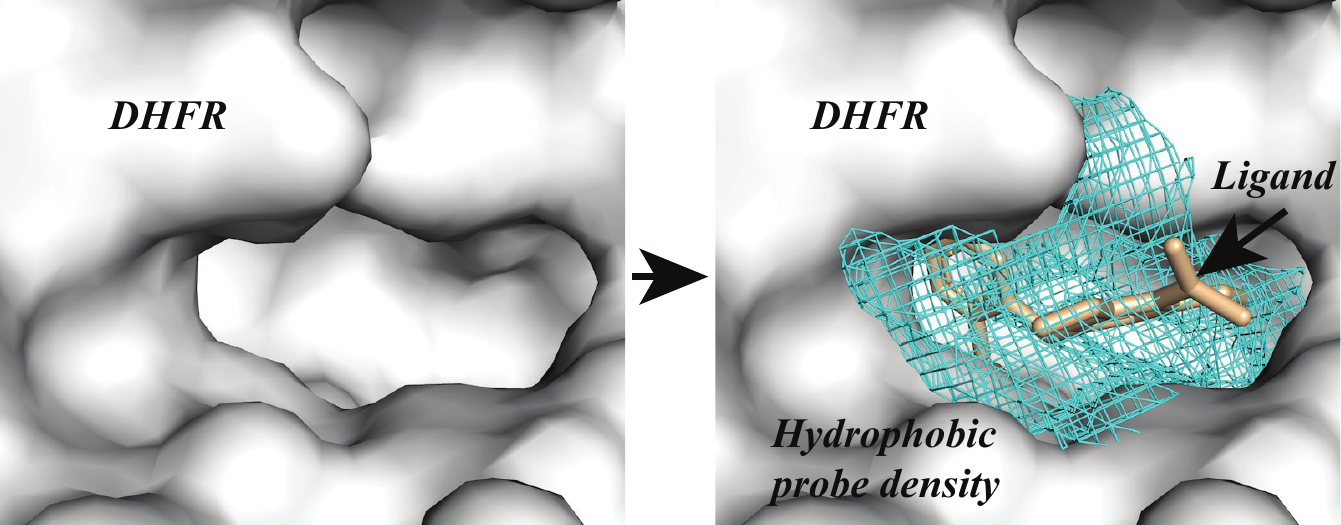}

\end{tocentry}

\begin{abstract}
We present an extension of the Poisson-Boltzmann model in which the solute of interest is immersed in an assembly of self-orienting Langevin water dipoles, anions, cations, and hydrophobic molecules, all of variable densities. Interactions between charges are controlled by electrostatics, while hydrophobic interactions are modeled with a Yukawa potential. We impose steric constraints by assuming that the system is represented on a cubic lattice. We also assume incompressibility, i.e. all sites of the lattice are occupied. This model, which we refer to as the Hydrophobic Dipolar Poisson Boltzmann Langevin (HDPBL) model, leads to a system of two equations whose solutions give the water dipole, salt, and hydrophobic molecule densities, all of them in the presence of the others in a self-consistent way. We use those to study the organization of the ions, co-solvent and solvent molecules around proteins. 
In particular, peaks of densities are expected to reveal, simultaneously, the presence of compatible binding sites of different kinds on a protein. 
We have tested and validated the ability of HDPBL to detect pockets in proteins  that bind to hydrophobic ligands, polar ligands and charged small probes as well as to characterize the environment of membrane proteins.
\end{abstract}


\section{1. Introduction}
\label{sec:intro}

Proteins are unique among biomolecules in that their function is modulated by a wide variery of small molecules that use different types of interactions in their binding sites. 
These binding sites can bind natural substrates, as observed for example in enzyme active sites, as well as in allosteric regulatory sites.
They are also the target of artificial or non natural ligands, developed by the pharmaceutical industry as drugs that can inhibit or activate protein function. 
The identification and characterization of those binding sites are therefore essential steps for understanding and controlling protein function by structure-based drug design, and as such have been central themes in structural biology.
Many experimental methods have been developed for this purpose, based on NMR \cite{Shuker:1996}, X-ray crystallography \cite{Mattos:1996}, as well as other techniques \cite{Hardy:2004}. 
Those techniques, however, are time-consuming and often expensive. 
This has led to parallel developments in computational structural biology with the same goals of identifying druggable binding sites in proteins \cite{Henrich:2010,
Konc:2014, Zhao:2020}. 
The corresponding methods are usually fast and easy to implement. Their success, however, is often mitigated by the fact that they rely on severe simplifications, such as ignoring solvent effects.
A successful computational technique for identifying drug-binding sites requires adequate and efficient sampling of the conformation of the ligand and
of the environment within the putative binding site, including the presence of ordered water molecules and salt.
Molecular dynamics (MD) simulations provides a rigourous framework to implement such sampling \cite{Iveta:2012, Feng:2018, Sledz:2018}.
Of particular interest, the incorporation of co-solvents to mimic the ligand in such simulations
improves the sampling and the identification of binding sites \cite{Basse:2010, Tan:2012,
Tan:2014, Kalenkiewicz:2015, Kimura:2017, Schmidt:2019}.
Molecular dynamics, however, is time-consuming as proper sampling requires long simulation times.
In this paper we propose a complementary technique based on statistical mechanics for identifying simultaneously 
one or several binding sites of different kinds in a given protein.

In an MD simulation interactions between atoms are usually described by semi-empirical ``force fields", 
with different levels of approximations (for recent reviews, see \cite{Nerenberg:2018, Huang:2018, Inakollu:2020, Spoel:2020}).
Applications of these force fields imply that the positions of all atoms be known.
While this seems to be a simple requirement, it is unfortunately difficult
to meet when modeling solvated large biomolecular systems.
This is mostly due to the inherent difficulties in accounting for the mobile 
solvent molecules and ions that surrounds the solute 
and of the size of the system that increases significantly when one includes solvent and ions in the simulation.
To circumvent these difficulties, there has been continuous efforts
to develop simplified models that remain physically accurate.
Many of these models consider the solvent implicitly, reducing the
solute-solvent interactions to their mean field characteristics.
In such models the solvent is treated as a dielectric continuum;
they are referred to as continuum dielectric models.
The Poisson-Boltzmann theory is one such model, unquestionably the most
popular,  which provides a framework for calculating the
electrostatics solvation free energy of a solute in such a dielectric continuum.
The corresponding PB equation (PBE) is not however the remedy to all problems associated
with characterizing the electrostatics interactions for a biomolecule:
it remains a mean field approximation, with known limitations.
Many improvements have been proposed to the PB equation.
The novel technique proposed in this paper is one such extension.
We first review the current theoretical developments
around the PB model and their relationships with our new model.

PBE is only a mean-field approximation to the multibody problem of
solvent-solute electrostatics interactions.
It is based on several approximations that have proved to be limitations in some cases.
For example, PBE does not include effects due to ion size or ion-ion correlations 
in its treatment (for review, see Grochowski and Trylska \cite{Grochowski:2007}).
Solutions have been proposed to account for at least ion size
using either a single size \cite{Borukhov:1997} or two different sizes \cite{Chu:2007},
yielding a size-modified Poisson-Boltzmann equation (SMPB), which is used to study
ion channels \cite{Xie:2020} and electrolyte solutions \cite{Stein:2019}.
Classical PBE handles solvent implicitly, 
with a fixed dielectric constant inside the biomolecule (usually set to 2-4)
that abruptly jumps to 80 at the interface between the biomolecule and the solvent.
This approximation leads to too much importance to the definition of this interface, 
usually set to the molecular surface or the vdW surface of the solute \cite{Pang:2013}. 
Nicholls and colleagues \cite{Grant:2001} proposed to solve this problem by 
using a Gaussian representation of the atoms of the solutes. Their solution, however,
does not circumvent an even more important limitation of the PB equation:
because of polarization effects in the vicinity of charges, 
a representation of the solvent as a homogeneous dielectric medium is bound
to be erroneous close to the interface.
We have developed an extension to the PB equation in which the solvent is described
as an assembly of interacting dipoles on a lattice gas to account for the non uniform
dielectric property of the solvent \cite{Azuara:2006, Abrashkin:2007, Azuara:2008, Koehl:2009, Koehl:2010}.
We referred to the corresponding equation as the Dipolar Poisson-Boltzmann-Langevin (DPBL) equations.
Here we describe an extension to DPBL, which we refer to as HDPBL, in which we allow 
for the addition of an hydrophobic cosolvent in the environment of the solute of interest.
These cosolvent molecules interact with each other and with hydrophobic ``charges" set on hydrophobic
groups inside the solute (usually CH2 and CH3 groups). Those interactions are represented with
a Yukawa potential. The electrostatic field and an additional attractive field associated with the Yukawa potential
are found to be solution of a system of two PB-like partial differential equations (PDE).  Those
fields are used to compute the water, ions, and co-solvent densities that are then used to map the
characteristics of pockets in the neighborhood of the protein of interest.

The HDPBL model proposed in this paper is an alternate approach to the cosolvent-enhanced MD simulations.
 Its novelty and possible advantages compared to these simulations are based on
\begin{itemize}
\item[i)] \emph{Improved sampling}.  
The conformational space and degrees of freedom of solvent, ions, and co-solvent molecules 
within and around the target biomolecule are sampled efficiently 
and their densities are computed self-consistently.
\item[ii)] \emph{Multiprobe exploration}. HDPBL enables incorporation of multiple probe types in the analyses.
The output of HDPBL are the densities of dipoles representing a polar solvent, anions, cations,
and hydrophobic cosolvent molecules as a function of position, thereby allowing for the identification
of polar, positively charged, negatively charged, and hydrophobic binding sites simultaneously.
\item[iii)] \emph{Efficiency}. MD simulations solve Newton's equations at the atomic level numerically
to construct a picture of the dynamics of the biomolecule of interest. Those simulations require significant
computing time to provide adequate sampling. In contrast, HDPBL relies on solving once a system of coupled
second order PDEs. We will show that this can be performed efficiently using standard techniques for solving elliptic PDEs.
\end{itemize}

This paper is organized as follows. The next section provides a complete overview of the HDPBL model.
The following section describes our implementation of a solver of HDPBL in a program we refer to as
AquaVit.  
AquaVit is a continuation of AquaSol \cite{Koehl:2010}; 
it is heavily based on the package MG developed by Michael Holst  \cite{Holst_thesis}.  
The next section provides examples of the usefulness of HDPBL for detecting and characterizing binding sites in proteins.
We conclude the paper with a discussion on the limitations of HDPBL, of possible improvements that
would circumvent those limitations, as well as on possible developments around HDPBL.

\section{2. The HDPBL model}
\label{sec:model}

\subsection{2.1 The system: solute, salt, solvent, and cosolvent}

We consider a fixed solute molecule (protein) in
a solution containing water, a $z:z$ salt (e.g. NaCl for which the valence $z$ is $1$), and a
cosolvent, i.e. neutral hydrophobic molecules. 
Those hydrophobic molecules will serve as probes to assess the hydrophobic
environment within, and around the solute of interest.
They may be considered as methane, xenon, krypton, or any other small hydrophobic molecule.

We assume that the system comprises
$N_{w}$ water molecules, $N_{s}$ salt molecules (that is $N_{s}$
cations of charge $+ze_c$ and $N_{s}$ anions of charge $-ze_c$ where
$e_c$ is the elementary electronic charge) and $N_{h}$ inert hydrophobic molecules. 
Those molecules are only present in the region outside of the solute, which
is characterized by its molecular surface. The water molecules
are modeled as dipoles with fixed dipole moment $p_{0}$. The charged
ions and the water dipoles interact through the Coulomb interaction,
and all particles are subject to steric repulsion. To model this steric
repulsion, we assume that the system is represented on a cubic lattice
gas of lattice spacing $a$ and that at each site of the lattice there
is either no particle, or one ion or one water molecule or one hydrophobic
molecule. Each mobile particle is thus modeled as a hard sphere with
radius $a/2$. We denote respectively by $n_{+}(\mathbf{r}),n_{-}(\mathbf{r}),n_w(\mathbf{r}),n_h(\mathbf{r})$
the occupation numbers of site $\mathbf{r}$ by a cation, an anion, a water
or a hydrophobic molecule, where each of these numbers are 1 or 0 depending
whether the corresponding particle is present or absent.
Note that $\mathbf{r}$ represents one of the cuboids of the cubic lattice.
the steric constraint imposes that there is at most one particle at
any site $\mathbf{r}$, 
\begin{equation}
n_{+}(\mathbf{r})+n_{-}(\mathbf{r})+n_w(\mathbf{r})+n_h(\mathbf{r})=0\:{\rm {or}}\:1
\label{eq:steric}
\end{equation}
If the system is incompressible, the above sum is strictly
equal to 1, meaning that there is exactly one particle at each site.
The relationships between the occupation numbers of all species and their
number of molecules are given by the additional constraints
\begin{eqnarray}
\sum_{\mathbf{r}}n_{+}(\mathbf{r}) & = & N_{s}\nonumber \\
\sum_{\mathbf{r}}n_{-}(\mathbf{r}) & = & N_{s}\nonumber \\
\sum_{\mathbf{r}}n_w(\mathbf{r}) & = & N_{w}\nonumber \\
\sum_{\mathbf{r}}n_h(\mathbf{r}) & = & N_{h}
\label{eq:constraints}
\end{eqnarray}
The corresponding particle densities are given by
\begin{eqnarray}
\rho_{\pm}(\mathbf{r}) & = & \frac{n_{\pm}(\mathbf{r})}{a^3} \nonumber \\
\rho_{w}(\mathbf{r}) & = & \frac{n_w(\mathbf{r})}{a^3} \nonumber \\
\rho_{h}(\mathbf{r}) & = & \frac{n_{h}(\mathbf{r})}{a^3} \nonumber \\
\end{eqnarray}
We emphasize that all the occupation numbers of mobile particles are zero inside the solute, since we assume that mobile particles cannot penetrate the solute.
Let $c_s$, $c_w$, and $c_h$ be the bulk concentrations of salt, water, and hydrophobic probes, respectively.
We introduce the volume fraction $\Phi$ for each type of particle
\begin{eqnarray}
\Phi_{s} & = & 2c_{s}a^{3} \nonumber \\
\Phi_{w} & = & c_{w}a^{3} \nonumber \\
\Phi_{h} & = & c_{h}a^{3}
\label{eqn:phi}
\end{eqnarray}
In addition, we consider a volume fraction for vacancies:
\begin{equation}
\Phi_{v}=1-(\Phi_{s}+\Phi_{w}+\Phi_{h})
\end{equation}
$\Phi_{v}$ should be understood as follows. If the system is incompressible, $\Phi_{v}=0$ and the concentrations of salt, water, and hydrophobic probes are necessarily dependent. Otherwise, $\Phi_v$ is positive, and vacancies are possible in the environment of the solute. In this case, $c_s$, $c_w$ and $c_h$ are independent, although they still need to satisfy $\Phi_s + \Phi_w + \Phi_h = 2c_{s}a^{3} + c_{w}a^{3} + c_{h}a^{3} \le 1$.

We denote by $v(\mathbf{r})=\frac{1}{4\pi\varepsilon_{0}r}$ the Coulomb potential,
where $\varepsilon_{0}$ is the dielectric permittivity of the vacuum and $r=|\mathbf{r}|$.

The hydrophobic interactions between the hydrophobic molecules are captured
by an attractive Yukawa potential

\begin{equation}
w(\mathbf{r})=-\frac{w_{0}}{4\pi} \frac{e^{-\kappa r}}{r}
\label{eq:Yukawa}
\end{equation}
where $\kappa=1/l_h$ defines the range of the hydrophobic interaction,
and $w_{0}>0$ its strength. The negative sign in Equation \ref{eq:Yukawa}
denotes the attractive nature of the interaction. 

\subsection{2.2 An effective free energy for the system}

The canonical partition function of the system on the lattice can be written as
\begin{eqnarray}
Z_{c} (N_s, N_w, N_h)  & = & \sum_{\{n_{\pm}(\mathbf{r})=0,1 \}} \sum_{\{n_w(\mathbf{r})=0,1\}} \sum_{\{n_h(\mathbf{r})=0,1\}}
\:\delta\left(\sum_{\mathbf{r}}n_{+}(\mathbf{r})-N_{s}\right)\delta\left(\sum_{\mathbf{r}}n_{-}(\mathbf{r})-N_{s}\right)\nonumber \\
 & \times & \delta\left(\sum_{\mathbf{r}} n_w(\mathbf{r})-N_{w}\right)\delta\left(\sum_{\mathbf{r}}n_h(\mathbf{r})-N_{h}\right)\nonumber \\
 & \times & 
\int\prod_{\mathbf{r}}d\mathbf{p}_w(\mathbf{r}) \delta\left( \mathbf{p}_w^2(\mathbf{r}) - p_0^2  \right) \nonumber \\
&&\exp\left(-\frac{\beta}{2}\sum_{\mathbf{r},\mathbf{r}'}\rho_{c}(\mathbf{r})v({\mathbf{r}} - {\mathbf{r}'} )\rho_{c}(\mathbf{r}')-\frac{\beta}{2}\sum_{\mathbf{r},\mathbf{r}'}\rho_H(\mathbf{r})w({\mathbf{r}} - {\mathbf{r}'} )\rho_H(\mathbf{r}')\right) \nonumber \\
\label{eq:partition}
\end{eqnarray}
where $\beta=1/k_{B}T$ with $T$ the temperature and $k_{B}$ the
Boltzmann constant. The notation $\sum_{\{n_{\pm}(\mathbf{ r})=0,1\}}$ denotes a sum over all possible combinations of values of the  occupation numbers (0 or 1) at all lattice sites. 
The total charge density $\rho_c$ in (\ref{eq:partition}) is the sum of the charge densities of the ions, of the point-like water dipoles
and of the charges of the solute
\begin{equation}
\rho_{c}(\mathbf{r})=ze_c(\rho_{+}(\mathbf{r})-\rho_{-}(\mathbf{r}))-\mathbf{p}_w(\mathbf{r})\cdot \nabla \rho_w(\mathbf{r})+\rho_{f}(\mathbf{r})
\label{eq:density}
\end{equation}
where $\rho_f(\mathbf{r})$ is the charge density of the fixed charges of the solute at point $\mathbf{r}$ and $\mathbf{p}_w(\mathbf{r})$ is the dipole moment of the water molecule at the same point. These dipole moments have a fixed magnitude $p_{0}$ but can take all possible orientations.
This is accounted for by the integral over all dipole orientations in (\ref{eq:partition}).
 \\
Finally, $\rho_{H}(\mathbf{r})$ is the sum of the density $\rho_h(\mathbf{r})$ of mobile hydrophobic particles  and of the density $\rho_p(\mathbf{r})$ of the hydrophobic sites of the fixed solute
\begin{equation}
\rho_{H}(\mathbf{r})=\rho_h(\mathbf{r}) + \rho_{p}(\mathbf{r})
\label{eq:density}
\end{equation}

Going to the grand canonical ensemble and introducing the chemical
potentials $\mu$ and fugacities $\lambda$ of the various species
\begin{eqnarray*}
\lambda_{s} & = & e^{\beta\mu_{s}}\\
\lambda_{w} & = & e^{\beta\mu_{w}}\\
\lambda_{h} & = & e^{\beta\mu_{h}}
\end{eqnarray*}
the grand partition function can be written as
\begin{eqnarray}
\Xi & = & \sum_{N_s=0}^{+\infty}\sum_{N_w=0}^{+\infty}\sum_{N_h=0}^{+\infty}
\: \frac{e^{2 \beta\mu_{s} N_s +\beta\mu_{w} N_w + \beta\mu_{h} N_h}}{ (N_s !)^2 N_w! N_h!} Z_{c} (N_s, N_w, N_h) 
 \label{eq:grandcanonical}
\end{eqnarray}
where the canonical partition function $Z_c$ is given in (\ref{eq:partition}).

Following the formalism introduced in \cite{Azuara:2006, Abrashkin:2007, Azuara:2008, Koehl:2009} we perform
two Hubbard-Stratonovich transforms within Equation \ref{eq:grandcanonical} and integrate
over the dipole moments. After a few standard manipulations, the partition
function can be written in an exact manner as an integral over two
fields $\varphi(\mathbf{r})$ and $\psi(\mathbf{r})$ corresponding to the electrostatic and the Yukawa interactions, respectively:
\begin{eqnarray}
\Xi & = & \int\mathcal{D}\varphi(\mathbf{r})\mathcal{D}\psi(\mathbf{r}) e^{-\left(\frac{\beta}{2}\int d\mathbf{r} d\mathbf{r}' \phi(\mathbf{r}) v^{-1}(\mathbf{r} - \mathbf{r}') \phi(\mathbf{r}') + \frac{\beta}{2}\int d\mathbf{r}\,d\mathbf{r}'\psi(\mathbf{r})w^{-1}({\mathbf{r}} - {\mathbf{r}'} )\psi(\mathbf{r}')\right)}\nonumber \\
 & \times & \exp{\left(-i\beta\int d\mathbf{r}\varphi(\mathbf{r})\rho_{f}(\mathbf{r})-\beta\int d\mathbf{r}\psi(\mathbf{r})\rho_{p}(\mathbf{r})\right)}\nonumber \\
 & \times & \prod_{\mathbf{r}}\left(\lambda_v +2\lambda_{s}\cosh\left(i\beta ze_c\varphi(\mathbf{r})\right)+\lambda_{w}\frac{\sinh\left(ip_{0}\beta\left|\nabla\varphi(\mathbf{r})\right|\right)}{ip_{0}\beta\left|\nabla\varphi(\mathbf{r})\right|}+\lambda_{h}e^{- \beta\psi(\mathbf{r})}\right)^{\gamma(\mathbf{r})}
 \label{eq:partfunc}
\end{eqnarray}
where we have introduce a pseudo fugacity for vacancies $\lambda_v$ such that $\lambda_v=0$ if the system is incompressible, and $\lambda_v=1$ otherwise, and $\gamma(\mathbf{r})$ is the indicator function for the points available
to the mobile particles, namely $\gamma(\mathbf{r})=1$ outside the solute, and $\gamma(\mathbf{r})=0$ inside.

The operator $w^{-1}(\mathbf{r})$ is the inverse of the Yukawa interaction
(\ref{eq:Yukawa}), and is given by
\begin{eqnarray*}
w^{-1}({\mathbf{r}} - {\mathbf{r}'} )=-\frac{1}{w_{0}}\left(-\nabla^{2}+\kappa^{2}\right)\delta({\mathbf{r}} - {\mathbf{r}'} )
\end{eqnarray*}

To simplify the notations and the equations, we  take the continuous limit $a \to 0$ in the lattice product above, and treat the fields $\varphi(\mathbf{r})$ and $\psi(\mathbf{r})$ as defined in the whole space. The sums are replaced by integrals according to
\begin{equation}
\sum_{\mathbf{r}} = \frac{1}{a^3} \int d \mathbf{r}
\end{equation}
where the integral on the right side is over the whole 3D space,
and we obtain
\begin{eqnarray}
 &  & \Xi =\int\mathcal{D}\varphi(\mathbf{r})\mathcal{D}\psi(\mathbf{r})e^{\frac{\beta\varepsilon_{0}}{2}\int d\mathbf{r}\left( \nabla\varphi(\mathbf{r})\right)^{2}-\frac{\beta}{2 w_{0}}\int d\mathbf{r} \left(\left(\nabla\psi(\mathbf{r})\right)^{2}+\kappa^{2}\psi^{2}(\mathbf{r})\right)}\nonumber \\
 &  & \times\exp{\left(-i\beta\int d\mathbf{r}\varphi(\mathbf{r})\rho_{f}(\mathbf{r})- \beta\int d\mathbf{r}\psi(\mathbf{r})\rho_{p}(\mathbf{r})\right)}\nonumber \\
 &  & \times\exp\left(\frac{1}{a^{3}}\int d\mathbf{r}\gamma(\mathbf{r})\ln\left(\lambda_v +2\lambda_{s}\cosh\left(i\beta ze_c\varphi(\mathbf{r})\right)+\lambda_{w}\frac{\sinh\left(ip_{0}\beta\left|\nabla\varphi(\mathbf{r})\right|\right)}{ip_{0}\beta\left|\nabla\varphi(\mathbf{r})\right|}+\lambda_{h}e^{- \beta\psi(\mathbf{r})}\right)\right) \nonumber \\
 \label{eq:funcpart}
\end{eqnarray}
The functional integral in Equation \ref{eq:funcpart} is evaluated by the Saddle-Point
Approximation. This method, which is also called Mean-Field Theory,
consists of minimizing the exponent in the above equation with respect
to the two fields $\varphi$ and $\psi$. The field $\varphi$ is pure imaginary at the saddle-point, while $\psi$ is real, and the exponent
above can be written as an effective free energy as
\begin{eqnarray}
\mathcal{F} & = & -\frac{\varepsilon_{0}}{2}\int d\mathbf{r}\left(\nabla\varphi(\mathbf{r})\right)^{2}+\frac{1}{2 w_{0}}\int d\mathbf{r}\left(\left(\nabla\psi(\mathbf{r})\right)^{2}+\kappa^{2}\psi^{2}(\mathbf{r})\right)\nonumber \\
 & + & \int d\mathbf{r}\varphi(\mathbf{r})\rho_{f}(\mathbf{r})+\int d\mathbf{r}\psi(\mathbf{r})\rho_{p}(\mathbf{r})\nonumber \\
 & - & \frac{k_{B}T}{a^{3}}\int d\mathbf{r}\gamma(\mathbf{r})\ln\left(\lambda_v +2\lambda_{s}\cosh\left(\beta ze_c\varphi(\mathbf{r})\right)+\lambda_{w}\frac{\sinh\left(p_{0}\beta\left|\nabla\varphi(\mathbf{r})\right|\right)}{p_{0}\beta\left|\nabla\varphi(\mathbf{r})\right|}+\lambda_{h}e^{-\beta\psi(\mathbf{r})}\right) \nonumber \\
 \label{eq:free}
\end{eqnarray}

\subsection{2.3 Optimizing the free energy}

The mean field equations are the Euler-Lagrange equations obtained
by minimizing \ref{eq:free} with respect to $\varphi$ and $\psi$
\begin{eqnarray}
-\varepsilon_{0}\nabla^{2}\varphi(\mathbf{r}) & = & \rho_{f}(\mathbf{r})-\frac{2\lambda_{s}}{a^{3}}ze_c\gamma(\mathbf{r})\frac{\sinh\left(\beta ze_c\varphi(\mathbf{r})\right)}{\mathcal{D}(\mathbf{r})}+\frac{p_{0}}{a^{3}}\lambda_{w}\gamma(\mathbf{r})\nabla \cdot \left(\frac{\mathbf{\nabla}\varphi(\mathbf{r})}{\left|\nabla\varphi(\mathbf{r})\right|}\frac{g\left(p_{0}\beta\left|\nabla\varphi(\mathbf{r})\right|\right)}{\mathcal{D}(\mathbf{r})}\right)\nonumber \\
\frac{1}{w_{0}}\left(-\nabla^{2}+\kappa^{2}\right)\psi(\mathbf{r}) & = & -\rho_{p}(\mathbf{r})-\frac{\lambda_{h}}{a^{3}}\gamma(\mathbf{r})\frac{e^{-\beta\psi(\mathbf{r})}}{\mathcal{D}(\mathbf{r})}\nonumber \\
\mathcal{D}(\mathbf{r}) & = & \lambda_v+2\lambda_{s}\cosh\left(\beta ze_c\varphi(\mathbf{r})\right)+\lambda_{w}\frac{\sinh\left(p_{0}\beta\left|\nabla\varphi(\mathbf{r})\right|\right)}{p_{0}\beta\left|\nabla\varphi(\mathbf{r})\right|}+\lambda_{h}e^{-\beta\psi(\mathbf{r})}
\label{eqn:meanfield}
\end{eqnarray}
where
\begin{eqnarray*}
g(x) & = & \frac{\cosh x}{x}-\frac{\sinh x}{x^{2}}
\end{eqnarray*}
Note that $\varphi(\mathbf{r}) \rightarrow 0$ and $\psi(\mathbf{r})\rightarrow \psi_0$ as $\mathbf{r}\rightarrow +\infty$, i.e. in the bulk, far from the solute. All coefficients in those equations are computed either from physical constants, or from input information describing the system, with the exception of the fugacities and $\psi_0$, which we derive now.

The fugacities are determined by the equations
\begin{eqnarray}
 -\lambda_{s}\frac{\partial(\beta F)}{\partial\lambda_{s}} &=& N_s \nonumber \\
 -\lambda_{w}\frac{\partial(\beta F)}{\partial\lambda_{w}} &=& N_w \nonumber \\
 -\lambda_{h}\frac{\partial(\beta F)}{\partial\lambda_{h}} &=& N_h\label{eq:fugue}
\end{eqnarray}
These equations translate into
\begin{eqnarray}
 \int d\mathbf{r}\frac{2\lambda_{s}}{\mathcal{D}{(\mathbf{r})}}{\cosh\left(\beta ze_c\varphi(\mathbf{r})\right)} &=& a^{3}N_{s} \nonumber \\
 \int d\mathbf{r}\frac{\lambda_{w}}{\mathcal{D}(\mathbf{r})}\frac{\sinh\left(p_{0}\beta\left|\nabla\varphi(\mathbf{r})\right|\right)}{p_{0}\beta\left|\nabla\varphi(\mathbf{r})\right|} &=& a^{3}N_{w} \nonumber \\
\int d\mathbf{r}\frac{\lambda_{h}}{\mathcal{D}(\mathbf{r})} e^{-\beta\psi(\mathbf{r})}&=& a^{3}N_{h}
\label{eq:fugue1}
\end{eqnarray}
Assuming that the volume of the solution is large compared to that of the solute,
\begin{eqnarray}
\frac{\lambda_{s}}{\lambda_v+2\lambda_{s}+\lambda_{w}+\lambda_{h}e^{-\beta\psi_{0}}} &=& a^{3}c_{s} = \frac{ \Phi_s}{2} \nonumber \\
 \frac{\lambda_{w}}{\lambda_v+2\lambda_{s}+\lambda_{w}+\lambda_{h}e^{-\beta\psi_{0}}} &=& a^{3}c_{w}  = \Phi_w \nonumber \\
 \frac{\lambda_{h}e^{-\beta\psi_{0}}}{\lambda_v +2\lambda_{s}+\lambda_{w}+\lambda_{h}e^{-\beta\psi_{0}}} &=& a^{3}c_{H} = \Phi_h
 \label{eqn:sysl}
\end{eqnarray}
where  $\Phi_s$, $\Phi_w$ and $\Phi_h$ the volume fractions defined in Equation \ref{eqn:phi}. Recall that $\Phi_v = 1.0 - \Phi_s - \Phi_w - \Phi_h$ is the molar fraction of vacancies.
We consider two cases:
\begin{itemize}
\item[i)] \emph{Compressible system}

In a compressible system, we allow for vacancies and $\Phi_v \ne 0$ and $\lambda_v = 1$. Solving the system in Equation \ref{eqn:sysl}, we get:
\begin{eqnarray}
2\lambda_{s} & = & \frac{\Phi_{s}}{\Phi_{v}}\nonumber \\
\lambda_{w} & = & \frac{\Phi_{w}}{\Phi_{v}}\nonumber \\
\lambda_{h}e^{-\beta\psi_{0}} & = & \frac{\Phi_{h}}{\Phi_{v}}
\label{eq:fractions}
\end{eqnarray}

\item[ii)] \emph{Incompressible system}

In the case of an incompressible system, there are no vacancies, and
$\Phi_{s}+\Phi_{w}+\Phi_{h}=1$. The fugacities are not independent,
and it is possible to chose for example $\lambda_{w}=1$. The fugacities
are then given by
\begin{eqnarray}
\lambda_{w} & = & 1\nonumber \\
2\lambda_{s} & = & \frac{\Phi_{s}}{\Phi_{w}}\nonumber \\
\lambda_{h}e^{-\beta\psi_{0}} & = & \frac{\Phi_{h}}{\Phi_{w}}\nonumber \\
\Phi_{w} & = & 1-\Phi_{s}-\Phi_{h}\label{eq:frac_incomp}
\end{eqnarray}
\end{itemize}

The bulk value $\psi_{0}$ is given by
\begin{equation}
\psi_{0}=-\frac{w_{0}}{\kappa^{2}}\frac{\Phi_{h}}{a^{3}}
\label{eq:bulkpsi}
\end{equation}

Using the values derived above for the fugacities, In all cases, the meanfield equations \ref{eqn:meanfield}
can be rewritten as 
\begin{subequations}
\label{eqn:newmeanfield} 
\begin{align}
-\varepsilon_{0}\nabla^{2}\varphi(\mathbf{r}) & =\rho_{f}(\mathbf{r})-\frac{\Phi_{s}}{a^{3}}ze_{c}\gamma(\mathbf{r})\frac{\sinh\left(\beta ze_{c}\varphi(\mathbf{r})\right)}{\mathcal{D}_{a}(\mathbf{r})}+\frac{p_{0}}{a^{3}}\Phi_{w}\gamma(\mathbf{r})\nabla \cdot \left(\frac{\mathbf{\nabla}\varphi(\mathbf{r})}{\left|\nabla\varphi(\mathbf{r})\right|}\frac{g\left(p_{0}\beta\left|\nabla\varphi(\mathbf{r})\right|\right)}{\mathcal{D}_{a}(\mathbf{r})}\right)\label{eqn:newmeanfield1}\\
\frac{1}{w_{0}}\left(-\nabla^{2}+\kappa^{2}\right)\psi(\mathbf{r}) & =-\rho_{p}(\mathbf{r})-\frac{\Phi_{h}}{a^{3}}\gamma(\mathbf{r})\frac{e^{-\beta(\psi(\mathbf{r})-\psi_{0})}}{\mathcal{D}_{a}(\mathbf{r})}\label{eqn:newmeanfield2}
\end{align}
\end{subequations}
where 
\begin{eqnarray}
\mathcal{D}_a(\mathbf{r}) & = & \Phi_v +\Phi_s \cosh\left(\beta ze_c\varphi(\mathbf{r})\right)+\Phi_{w}\frac{\sinh\left(p_{0}\beta\left|\nabla\varphi(\mathbf{r})\right|\right)}{p_{0}\beta\left|\nabla\varphi(\mathbf{r})\right|}+\Psi_{H}e^{-\beta\psi(\mathbf{r})}
\end{eqnarray}
Note that $\mathcal{D}_{a}(\mathbf{r}) \rightarrow 1$ when $\mathbf{r} \rightarrow +\infty$.

\subsection{2.4 The HDPBL system of equations}

The meanfield equations \ref{eqn:newmeanfield} given above fully describe the system under study. 
Equation \ref{eqn:newmeanfield1} is a Dipolar Poisson-Boltzmann
Langevin (DPBL) equation \cite{Azuara:2006, Azuara:2008, Koehl:2010}, while equation \ref{eqn:newmeanfield2} is
a Poisson-Boltzmann-like equation that relates to the hydrophobic interactions involving the hydrophobic probes in the solvent and the hydrophobic charges on the solute. As a consequence, we refer to this system of equations as the Hydrophobic Dipolar Poisson-Boltzmann Langevin equations, or HDPBL equations in short. In the following, we provide modified
equations involving dimensionless potentials that are more amenable to a numerical solution, and we 
derive all constants necessary for those equations.

\subsubsection{2.4.1 Revisiting the DPBL-like equation \ref{eqn:newmeanfield1}}

The electrostatic potential $\varphi(\mathbf{r})$ and the field $\psi(\mathbf{r})$
are expressed in volts and Joules in the SI unit system, respectively.
It is common to consider instead the dimensionless potentials $u(\mathbf{r})$
and $v(\mathbf{r})$ defined as: 
\begin{subequations}
\label{eqn:dim} 
\begin{align}
u(\mathbf{r}) & =\frac{e_{c}\varphi(\mathbf{r})}{k_{B}T}=\beta e_{c}\varphi(\mathbf{r})\label{eqn:dim1}\\
v(\mathbf{r}) & =\frac{\psi(\mathbf{r})-\psi_{0}}{k_{B}T}=\beta(\psi(\mathbf{r})-\psi_{0})\label{eqn:dim2}\\
\end{align}
\end{subequations}
Equation \ref{eqn:newmeanfield1} can then be rewritten as 
\begin{eqnarray}
-\varepsilon_{0}\nabla^{2}u(\mathbf{r}) & =\beta e_{c}^{2}\rho_{fd}(\mathbf{r})-\frac{\Phi_{s}}{a^{3}}\beta ze_{c}^{2}\gamma(\mathbf{r})
\frac{\sinh\left(zu(\mathbf{r})\right)}{\mathcal{D}_{1}(\mathbf{r})}
+\frac{\beta e_{c}^2p_{e}}{a^{3}}\Phi_{w}\gamma(\mathbf{r})\nabla \cdot \left(\frac{\mathbf{\nabla} u(\mathbf{r})}{\left|\nabla u(\mathbf{r})\right|}\frac{g\left(p_{e}\left|\nabla u(\mathbf{r})\right|\right)}{\mathcal{D}_{1}(\mathbf{r})}\right)\label{eqn:eq1}
\end{eqnarray}
where we have defined $p_e = p_0/e_c$, and
\begin{eqnarray}
\mathcal{D}_{1}(\mathbf{r}) & = & \Phi_{v}+\Phi_{s}\cosh\left(zu(\mathbf{r})\right)+\Phi_{w}\frac{\sinh\left(p_{e}\left|\nabla u(\mathbf{r})\right|\right)}{p_{e}\left|\nabla u(\mathbf{r})\right|}+\Phi_{h}e^{-v(\mathbf{r})}\label{eqn:D1}
\end{eqnarray}
In equation \ref{eqn:eq1}, $\rho_{fd}$ is the density of fixed charges
expressed as fraction of electrons, hence the $e_{c}^{2}$ as a coefficient,
where the first $e_{c}$ comes from the change to a dimensionless potential,
and the second $e_{c}$ is factored from the solute charges.

Let us introduce the function $F(x)=\frac{g(x)}{x}$. Note that 
\begin{eqnarray}
F(x)=\frac{\cosh(x)}{x^{2}}-\frac{\sinh(x)}{x^{3}}=\frac{\sinh(x)}{x^{2}}\mathcal{L}(x)
\end{eqnarray}
where $\mathcal{L}(x)=\frac{1}{\tanh(x)}-\frac{1}{x}$ is the Langevin function and $F(x) \rightarrow 1/3$ when $x \rightarrow 0$..

After a few standard manipulations, taking into account that $\nabla \gamma(\mathbf{r})=\mathbf{0}$, equation \ref{eqn:eq1} can be
rewritten as: 
\begin{eqnarray}
-\nabla \cdot \left( \left(1+\gamma(\mathbf{r})C_{w}\Phi_{w}\frac{F\left(p_{e}\left|\nabla u(\mathbf{r})\right|\right)}{\mathcal{D}_{1}(\mathbf{r})} \right) \nabla u(\mathbf{r})\right) + \gamma(\mathbf{r})C_{s}\Phi_{s}\frac{\sinh\left(zu(\mathbf{r})\right)}{\mathcal{D}_{1}(\mathbf{r})} & =C_{f}\rho_{fd}(\mathbf{r})\label{eqn:neweq1}
\end{eqnarray}

where we have introduced the three constants: 
\begin{subequations}
\label{eqn:const} 
\begin{align}
C_{w} & =\frac{4 \pi l_B p_{e}^{2}}{a^{3}}\label{eqn:const2a}\\
C_{s} & =\frac{z 4\pi l_B}{a^{3}}\label{eqn:const2b}\\
C_{f} & =4\pi l_B \label{eqn:const2c}
\end{align}
\end{subequations}
where $l_B$ is the Bjerrum length in vacuum, namely
\begin{eqnarray}
l_B = \frac{\beta e_c^2}{4\pi \varepsilon_0}
\end{eqnarray}

As written, equation \ref{eqn:neweq1} is a PB-like second
order differential equation, with a relative permittivity $\varepsilon(\mathbf{r}, u, v)$
defined as: 
\begin{eqnarray}
\varepsilon(\mathbf{r},u,v)=1+\gamma(\mathbf{r})C_{w}\Phi_{w}\frac{F\left(p_{e}\left|\nabla u(\mathbf{r})\right|\right)}{\mathcal{D}_{1}(\mathbf{r})}\label{eqn:eps}
\end{eqnarray}
$\varepsilon(\mathbf{r},u,v)$ is 1 inside the solute, and depends on both the
position $\mathbf{r}$ and on the values of the potentials $u(\mathbf{r})$ and $v(\mathbf{r})$
at that position, for $\mathbf{r}$ outside the molecule. Equation \ref{eqn:eps}
gives a self-consistent representation of the dielectric permittivity
of the system. 

\subsubsection{2.4.2 Revisiting the PB-like equation \ref{eqn:newmeanfield2}}

Using the dimensionless potentials $u(\mathbf{r})$ and $v(\mathbf{r})$
defined in equations \ref{eqn:dim}, equation \ref{eqn:newmeanfield2}
becomes 
\begin{eqnarray}
\frac{1}{w_{0}}\left(-\nabla^{2}+\kappa^{2}\right)(v(\mathbf{r})+v_{0}) & =-\beta\rho_{p}(\mathbf{r})-\frac{\beta\Phi_{h}}{a^{3}}\gamma(\mathbf{r})\frac{e^{-v(\mathbf{r})}}{\mathcal{D}_{1}(\mathbf{r})}\label{eqn:eq2}
\end{eqnarray}
where 
\begin{eqnarray}
v_{0}=\beta\psi_{0}=-\frac{\beta w_0}{\kappa^2}\frac{\Phi_{h}}{a^{3}}
\end{eqnarray}
Note that $\beta w_0$ is a length, which we write as $l_Y$, and
\begin{eqnarray}
v_0 = \frac{l_Y}{\kappa^2}\frac{\Phi_{h}}{a^{3}}
\end{eqnarray}
We rewrite Equation \ref{eqn:eq2} as: 
\begin{eqnarray}
-\nabla^{2}v(\mathbf{r})+\kappa^{2}(v(\mathbf{r})+v_{0})+\gamma(\mathbf{r})\frac{l_{Y}}{a^{3}}\Phi_{h}\frac{e^{-v(\mathbf{r})}}{\mathcal{D}_{1}(\mathbf{r})}=-l_{Y}\rho_{p}(\mathbf{r})\label{eqn:neweq2}
\end{eqnarray}

\subsection{2.5 The densities or water dipoles, ions, and hydrophobic probes}

Once the dimensionless fields $u^{MF}(\mathbf{r})$ and $v^{MF}(\mathbf{r})$ have been derived as solutions of the HDPBL system of equations, the densities of the various molecules are given by

\begin{itemize}
\item[i)] Anions and cations:
\begin{eqnarray}
\rho_{\pm}(\mathbf{r}) & = & \frac{1}{a^{3}}\frac{\Phi_{s}e^{\mp z u^{MF}(\mathbf{r})}}{2\mathcal{D}_{1}^{MF}(\mathbf{r})}
\end{eqnarray}
\item[ii)] Salt:
\begin{eqnarray}
\rho_{s}(\mathbf{r}) & = & \frac{1}{a^{3}}\frac{\Phi_{s}\cosh(z u^{MF}(\mathbf{r}))}{\mathcal{D}_{1}^{MF}(\mathbf{r})}
\end{eqnarray}
\item[iii)] Water dipoles
\begin{eqnarray}
\rho_{w}(\mathbf{r}) & = & \frac{1}{a^{3}}\frac{\Phi_{w}}{\mathcal{D}_{1}^{MF}(\mathbf{r})}\frac{\sinh\left(p_{e}\left|\nabla u^{MF}(\mathbf{r})\right|\right)}{p_{e}\left|\nabla u(\mathbf{r})\right|}
\label{eqn:rhow}
\end{eqnarray}
\item[iv)] Hydrophobic probes
\begin{eqnarray}
\rho_{h}(\mathbf{r}) & = & \frac{1}{a^{3}}\frac{\Phi_{h}}{\mathcal{D}_{1}^{MF}(\mathbf{r})}e^{-v(\mathbf{r})}
\end{eqnarray}
\end{itemize}
where $\mathcal{D}_{1}^{MF}(\mathbf{r})=\Phi_{v}+\Phi_{s}\cosh\left(zu^{MF}(\mathbf{r})\right)+\Phi_{w}\frac{\sinh\left(p_{e}\left|\nabla u^{MF}(\mathbf{r})\right|\right)}{p_{e}\left|\nabla u^{MF}(\mathbf{r})\right|}+\Phi_{h}e^{-v^{MF}(\mathbf{r})}$.

Equation \ref{eqn:rhow} defines the local densities of water around the solute. In parallel, we can also compute the polarisation density

\begin{eqnarray}
\mathbf{P}(\mathbf{r}) & = & \frac{p_{0}}{a^{3}}\Phi_{w} \frac{g\left(p_{0} \beta \left| \mathbf{E}^{MF}(\mathbf{r} \right| \right)}{\mathcal{D}_{1}^{MF}(\mathbf{r} )}  \hat {\mathbf{E}}^{MF}(\mathbf{r})
\end{eqnarray}
where $\hat {\mathbf{E}}$ is the unit vector of the electric field $\mathbf{E}$. Using the expression for the water density, we have
\begin{equation}
\mathbf{P}(\mathbf{r} ) = \rho_w(\mathbf{r} ) {\mathcal L}\left( p_0 \beta \left| \mathbf{E}^{MF}(\mathbf{r}) \right| \right) \hat {\mathbf{E}}^{MF}(\mathbf{r})
\label{eqn:polar2}
\end{equation}
where $\mathcal{L}$ is the usual Langevin function (see above). For small electric field $\mathbf{E}$, eq. \ref{eqn:polar2} becomes the standard linear relation
\begin{equation}
\mathbf{P}(\mathbf{ r}) = \frac{p_0^2 \beta c_w}{3} \mathbf{E}^{MF}(\mathbf{r})= \alpha \mathbf{E}^{MF}(\mathbf{r})
\end{equation}
Note that if we use the expression for the polarisation density $\mathbf{P}(\mathbf{r})$ we can rewrite the DPBL-like equation \ref{eqn:neweq1} as
\begin{equation}
\label{polar}
\nabla \cdot (\varepsilon_0 \mathbf{E}(\mathbf{r}) + \mathbf{P}(\mathbf{r} )) = \rho_f(\mathbf{r}) + \rho_{ions}(\mathbf{r}) 
\end{equation}

\section{3. Numerical solutions to the HDPBL system of equations}
\label{sec:comput}

Let us first recall the system of equations HDPBL:
\begin{subequations}
\label{eqn:newhdpbl} 
\begin{align}
-\nabla\cdot \left( \varepsilon(\mathbf{r},u,v) \nabla u(\mathbf{r}) \right)+\gamma(\mathbf{r})C_{s}\Phi_{s}\frac{\sinh\left(zu(\mathbf{r})\right)}{\mathcal{D}_{1}(\mathbf{r},u,v)} & =C_{f}\rho_{fd}(\mathbf{r})\label{eqn:newhdpbl1}\\
-\nabla^{2}v(\mathbf{r})+\kappa^{2}(v(\mathbf{r})+v_{0})+\gamma(\mathbf{r})\frac{l_{Y}}{a^{3}}\Phi_{h}\frac{e^{-v(\mathbf{r})}}{\mathcal{D}_{1}(\mathbf{r},u,v)} &=-l_{Y}\rho_{p}(\mathbf{r})\label{eqn:newhdpbl2}
\end{align}
\end{subequations}
where $\mathcal{D}_1(\mathbf{r},u,v)$ and $\varepsilon(\mathbf{r},u,v)$ are functions of the position, $\mathbf{r}$, and of the fields $u$ and $v$:
\begin{eqnarray}
\varepsilon(\mathbf{r},u,v) &=& 1+\gamma(\mathbf{r})C_{w}\Phi_{w}\frac{F\left(p_{e}\left|\nabla u(\mathbf{r})\right|\right)}{\mathcal{D}_{1}(\mathbf{r},u,v)} \nonumber \\
\mathcal{D}_{1}(\mathbf{r},u,v) & = & \Phi_{v}+\Phi_{s}\cosh\left(zu(\mathbf{r})\right)+\Phi_{w}\frac{\sinh\left(p_{e}\left|\nabla u(\mathbf{r})\right|\right)}{p_{e}\left|\nabla u(\mathbf{r})\right|}+\Phi_{h}e^{-v(\mathbf{r})}\label{eqn:newD1}
\end{eqnarray}
The two equations in this system are dependent as they both involve the two fields $u$ and $v$. They are PB-like, but cannot be solved directly using a PB solver. Equation \ref{eqn:newhdpbl} for example is a second order elliptic non linear PDE, like PB; however, its response coefficients (the coefficients in the divergence term) are not constant, as they are non linear functions of the  two fields $u$ and $v$. For equation \ref{eqn:newhdpbl}, it is the Helmholtz term that is a non linear function of the  two fields $u$ and $v$. Instead of considering a new specific solver for each of those equations, we propose to use a standard inexact Newton method developed for the PB equation by Hold and Saied \cite{Holst:1995} within a self-consistent algorithm for solving the HDPBL system, as described in algorithm \ref{alg:algorithm1}. 

\begin{algorithm}
\caption{\emph{AquaVit: Self Consistent Newton method for solving the DPBL equation}}
\label{alg:algorithm1}
\begin{algorithmic}
\STATE	Initialize	$u_0(\mathbf{r}) = 0$ and $v_0(\mathbf{r}) = 0$, $\forall \mathbf{r}$ \\
\FOR {$n=0,\ldots$ until convergence}
\\\hrulefill
\STATE     (1)    Solve equation \ref{eqn:newhdpbl1} for $u(\mathbf{r})$ with $v(\mathbf{r})$ fixed
\\\hrulefill
\STATE (1a) Initialize  a field $\psi(\mathbf{r})=0, \forall \mathbf{r}$; \\
\quad \quad \quad Define $F(\mathbf{r}, \psi) = - \nabla \cdot \left( \varepsilon(\mathbf{r},\psi, v_n)  \nabla\psi(\mathbf{r}) \right)
+\gamma(\mathbf{r})C_{s}\Phi_{s}\frac{\sinh\left(z \psi(\mathbf{r})\right)}{\mathcal{D}_{1}(\mathbf{r},\psi,v_n)} - C_{f}\rho_{fd}(\mathbf{r})$
\FOR {$m=0,\ldots$ until convergence}
\STATE	(1b)	Set $\varepsilon_m(\mathbf{r}) = \varepsilon(\mathbf{r}, \psi_{m},v_{n})$ 
\STATE	(1c)  Set $\mathcal{D}_m(\mathbf{r})=\mathcal{D}_1(\mathbf{r},\psi_{m},v_{n})$ \\
                            \quad \quad Define $\displaystyle H_m(\mathbf{r},\psi) = \gamma(\mathbf{r})C_{s}\Phi_{s}\frac{\sinh\left(z\psi(\mathbf{r})\right)}{\mathcal{D}_{m}(\mathbf{r})}$
\STATE	(1d)    Solve the PB-like PDE: \\ 
		\quad \quad $\nabla \cdot \left( \varepsilon_n(\mathbf{r}) \nabla\psi(\mathbf{r}) \right) + H_n(\mathbf{r},\psi(\mathbf{r})) = C_{f}\rho_{fd}(\mathbf{r})$ \\ 
		\quad \quad for $\psi_{sol}$, using the inexact Newton method of Holst and Saied \cite{Holst:1995} 
\STATE	(1e)	Update $\psi$: \\
			\quad \quad $\psi_{m+1}(\mathbf{r}) = \lambda \psi_{sol}(\mathbf{r}) + (1-\lambda) \psi_m(\mathbf{r}), \quad \forall \mathbf{r}$
\STATE	(1f)	Check for convergence:  if $\displaystyle \frac{\sum_{\mathbf{r}} \| \mathbf{F}(\mathbf{r},\psi_{m+1}\|}{\sum_{\mathbf{r}}\|\mathbf{F}(\mathbf{r},\psi_{0})\|} < TOL$, stop
\ENDFOR
\STATE(1g) Update $u_{n+1}(\mathbf{r}) =\psi_{m+1}(\mathbf{r}), \quad \forall \mathbf{r}$
\\\hrulefill
\STATE     (2)    Solve equation \ref{eqn:newhdpbl2} for $v(\mathbf{r})$ with $u(\mathbf{r})$ fixed
\\\hrulefill
\STATE    (2a) Initialize  a field $\phi(\mathbf{r})=0, \forall \mathbf{r}$;
\STATE	(2b)  Define $\displaystyle G(\mathbf{r},\phi) = \kappa^{2}(\phi(\mathbf{r})+v_{0})+\gamma(\mathbf{r})\frac{l_{Y}}{a^{3}}\Phi_{h}\frac{e^{-\phi(\mathbf{r})}}{\mathcal{D}_{1}(\mathbf{r},u_{n+1},\phi)}$
\STATE	(2c)    Solve the PB-like PDE: \\ 
		\quad \quad $-\nabla^{2}v(\mathbf{r}) + G(\mathbf{r},\phi(\mathbf{r})) = -l_{Y}\rho_{p}(\mathbf{r})$ \\ 
		\quad \quad for $\phi_{sol}$, using the inexact Newton method of Holst and Saied \cite{Holst:1995} 
\STATE(2d) Update $v_{n+1}(\mathbf{r}) = \lambda \phi_{sol}(\mathbf{r}) + (1-\lambda) v_n(\mathbf{r}), \quad \forall \mathbf{r}$
\STATE (3) Check for convergence:  if $\displaystyle \sum_{\mathbf{r}}\left| v_{n+1}(\mathbf{r}) - v_n(\mathbf{r}) \right| < TOL$, stop
\ENDFOR
\end{algorithmic}
\end{algorithm}

This algorithm alternatively solves for $u$ (part 1), given $v$, and then solves for $v$ (part 2), given $u$, until convergence, i.e. until those fields do not change anymore.
To solve for $u$, step (1b)  sets the diffusion coefficient $\varepsilon_n$ independent of the fields $u$ and $v$. Similarly, step (1c) defines a Helmholtz-like term $H_m$ whose value at position $\mathbf{r}$ only depends on the value of the potential at that position.  The PDE in step (1d) is then a PB equation that can be solved directly with a PB algorithm without modification. The updates in step (1e) and (2d) follow a typical trick for self-consistent methods that removes oscillations in the  convergence behavior.  
Note that here is no need to solve the PDE in step (1d) exactly. As its solution $\psi(\mathbf{r})$ is used as an correction  for the solution of the DPBL equation (step 1e), it is appropriate to use an approximation: the number of total iterations may then be higher, but this is compensated by the fact that the amount of work per iteration is smaller.

The algorithm described above was implemented in a software package, AquaVit. AquaVit is written in Fortran and is designed specifically to solve the HDPBL system of equations.
AquaVit is mostly inspired from AquaSol \cite{Koehl:2010}, and uses many routines from the Fortran package MG developed by Michael Holst  \cite{Holst_thesis}.
More information on the implementation is available in the paper describing Aquasol in length \cite{Koehl:2010}.

\section{4. Methods}
\label{sec:method}

\subsection{4.1 System set up}

The coordinates of the atoms of the solute(s) as well as their partial charges are read from a single file under the PQR format used by APBS. For large biomolecules, PQR files can be readily generated from the correspondent PDB \cite{Berstein:2000} files using the service PDB2PQR \cite{Dolinsky:2004}.
For all examples described below, we used the PARSE parameter dataset \cite{Sitkoff:1994} to assign charges. The PQR file may contain several molecules. The PQR file was then modified to add hydrophobic charges to all atoms. While those are not charge per se, they enable interactions between the hydrophobic probe in the solvent and the solute. Using the nomenclature of CHARMM, all atoms identified as CT (aliphatic carbon), CH1E (extended carbon, with 1 hydrogen), CH2E (extended carbon, with 2 hydrogens), CH3E (extended carbon, with 3 hydrogens), S (sulfur) and SH1E (extended atom S with one hydrogen) were assigned a non-zero hydrophobic charge of +2, while all other atoms had no non polar charges.

AquaVit starts by building a regular mesh  around the solutes.  The mesh is positioned such that its center matches with the center of the solute. The user provides the number of points and the mesh spacing in each directions. 
AquaVit checks that there is at least a distance of $2l_B^w$, ($l_B^w$ being the Bjerrum length in water at 300K, i.e. approximately 7 \AA) from any point on the surface of the solute to the closest edge of the mesh; if this condition is not met, the mesh size is adjusted accordingly.

The interface between the interior and exterior of the solutes is modeled based on their molecular surface. 
The molecular surface is the lower envelope obtained by rolling a water probe of radius $R_{probe}$ on the vdW surface of the molecule. A full description of how this surface is computed can be found in Ref. \cite{Koehl:2010}.

Classical treatment of electrostatics assigns a point charge to each atom, usually located at the center of the sphere representing this atom. 
The mesh considered in AquaVit is Cartesian; as such, the centers of the atoms of the solute(s) will most likely not coincide with its vertices.
We therefore need to project the atomic charges on the vertices of the mesh; we have used trilinear interpolation.

\subsection{4.2 Parameterizing the HDPBL system}

The HDPBL system of equations include 10 parameters: the number of vertices in the mesh in each dimension, the lattice size $a$, the temperature $T$, the concentrations of water $c_w$, salt, $c_s$, and hydrophobic probes, $c_h$, the valence $z$ of the anions and cations from the salt, the strength of the water dipole, $p_0$, and the parameters of the Yukawa potential, $l_Y (=\beta w_0)$ and $l_h (= 1/\kappa)$.

The mesh was set with 193 vertices in each direction, $x$, $y$, and $z$. Those vertices are equally spaced, and the distance between two vertex is computed automatically based on the size of the solute and the fact that the borders of the mesh are set to be at least $2l_B^w$ away from the solute.

Assuming incompressibility, in the presence of pure water, we expect $\Phi_w = 1$, i.e. that $c_w a^3 = 1$, where $c_w$ is the concentration of bulk water, namely 55 M. This leads to $a = 3.11$. In all the simulations described in the result section, we have considered monovalent (i.e. $z=1$) salt at 0.2 M and hydrophobic probes at 1 M. This amount of hydrophobic probes is equivalent to the amount used in ligand mapping molecular dynamics programs such as SILCS \cite{Guvench:2009} and SWISH \cite{Oleinikovas:2016}. Again, assuming incompressibility, the concentration of water is then fixed as we have the relation (see above):
\begin{eqnarray*}
2 c_s a^3 + c_w a^3 + c_h a^3 = 1
\end{eqnarray*}
Using the prescribed concentrations of salt and hydrophobic probes, and the lattice size $a=3.11$, this gives us an apparent concentration of water $c_w = 53.6$ M.

The parameter $l_h$ defines the range of the Yukawa potential. We have set it equal to the size of the lattice, i.e. $l_h=3.1$ \AA. $l_Y$ is a characteristic length for the Yukawa potential that directly relate to its strength. We have set it to $l_Y = 4$ \AA.

The temperature is set to 300K.

The experimental dipole moment of water is $1.85D$ in the gas phase. We have observed previously that with this value for $p_0$, we could not  obtain the correct dielectric permittivity of water in the DPBL model \cite{Azuara:2008}. As HDPBL is based on DPBL, we follow the recommendation of increasing $p_0$ which we set at $2.8D$.

In Table \ref{table:value1}, we give the values of the three constants $C_{w}$, $C_{s}$, $C_{f}$, and $p_e$ for a typical run of AquaVit with
$T=300K$ and the values of the different parameters given above.

\begin{table}[!hbt]
\centering \begin{threeparttable} \caption{Typical values for the constants in equation \ref{eqn:neweq1}}
\label{table:value1} 
%
\begin{tabular}{lp{3cm}p{3cm}p{4cm}}
\hline 
Name  & Expression & value\tnote{(a)}  & unit\tabularnewline
\hline 
$l_B$ & $\frac{\beta e_c^2}{4\pi \varepsilon_0}$ & 556.99995 & \AA \tabularnewline
$C_{w}$  & $\frac{4 \pi l_B p_{e}^{2}}{a^{3}}$ & $78.51664$  & Dimensionless \tabularnewline
$C_{s}$  & $\frac{z 4\pi l_B}{a^{3}} $ & $231.83464$  & \AA$^{-2}$ \tabularnewline
$C_{f}$  & $4\pi l_B$  & $6999.46779 $  & \AA \tabularnewline
$p_e$ & $\frac{p_0}{e_c}$ & 0.58195 & \AA \tabularnewline
\hline 
\end{tabular}\begin{tablenotes} 

\item (a) {\small{}$T=300K$, $a=3.11$ \AA\ (size of the lattice),
$p_{D}=2.8D$ (dipole moment of water), and $z=1$ (valence of salt
ions).} \end{tablenotes} \end{threeparttable} 
\end{table}
  
\subsection{4.2 Output format}

The output of AquaVit are the maps corresponding to the densities of the different species in the solvent, namely the water dipoles, the anions, cations, and the hydrophobic probes. While those maps are initially expressed as relative densities with respect to the bulk concentrations of the respective species, we have re-expressed those maps as $Z-$maps, using:
\begin{eqnarray*}
Z(\mathbf{r}) = \frac { \rho(\mathbf{r}) - \mu_{\rho}}{\sigma_{\rho}}
\end{eqnarray*}
where $\mu_{\rho}$ and $\sigma_{\rho}$ are the mean and standard deviation of the densities computed over all positions $\mathbf{r}$ outside of the solute, respectively.

\section{5. Results and Discussion}
\label{sec:results}

HDPBL differs from a standard PB model, as we have included hydrophobic probes around the solute of interest. Our primary goal is to use those probes to detect hydrophobic pockets in this solute. We have therefore tested the ability of HDPBL to detect pockets that bind to hydrophobic ligands as well as to characterize the environment of membrane proteins. HDPBL also provides the densities of anions, cations, and water dipole around the solute. We have included a test on detecting binding sites characterized with electrostatics interactions to illustrate their usefulness. Finally, we will describe limitations of HDBPL when it comes to detecting cryptic binding sites.

\subsection{5.1 Detecting hydrophobic pockets in proteins}

\begin{figure}[!tb]
\centering
\includegraphics[width=0.8\textwidth]{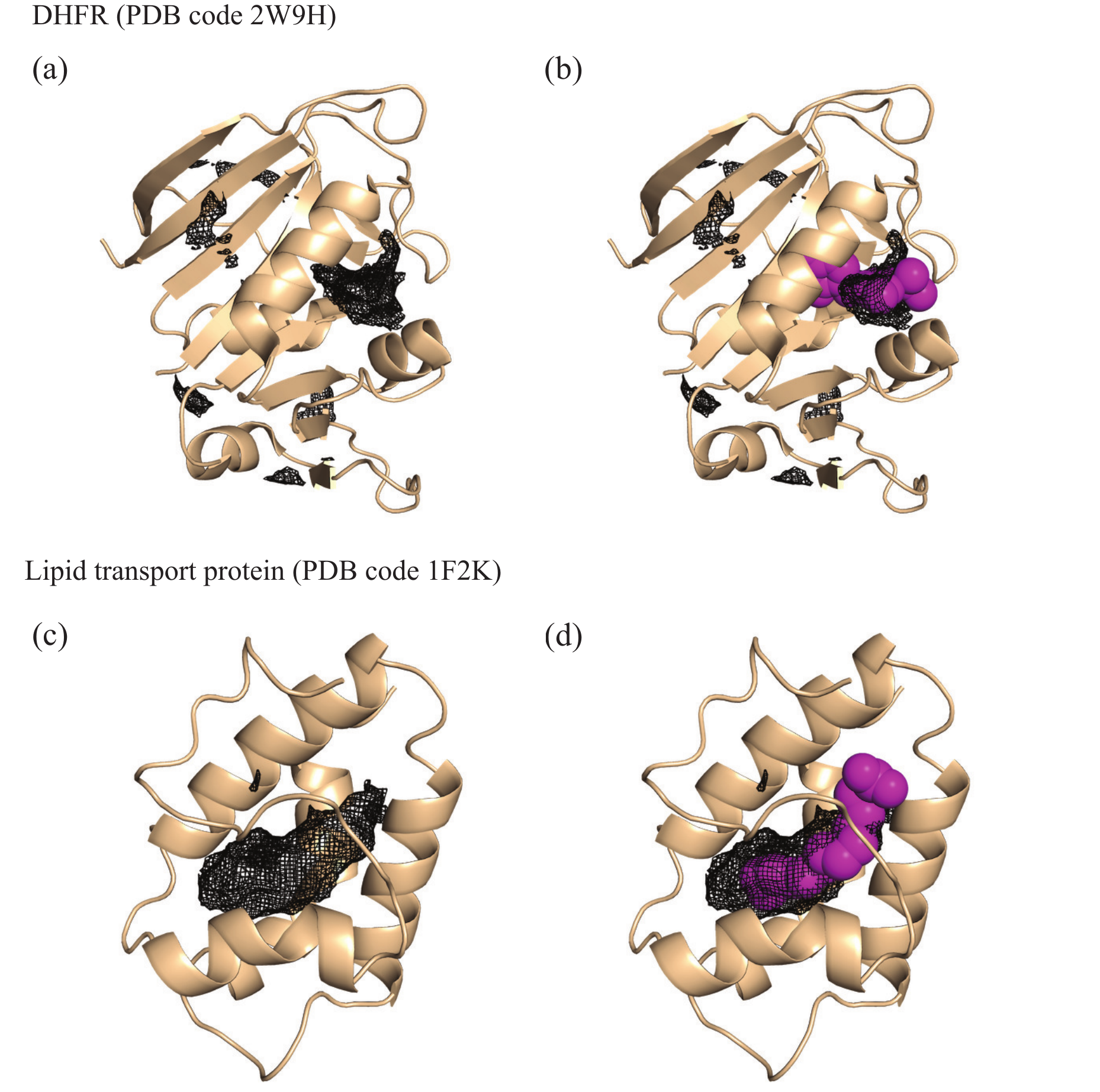}
	\caption{Hydrophobic occupancy maps, black mesh, superimposed onto the PDB structures for DHFR (PDB code 2W9H) and a lipid binding protein (LBP, PDB code 1FK2). The maps are derived from the
	densities of hydrophobic probes computed by AquaVit, and represented at +20 $\sigma$. Those maps are derived from the apo structure of the protein, i.e. in the absence of all crystallographic ligands and water molecules. In panels (a) and (c) we show the hydrophobic maps for DHFR and LBP superposed to the apo PDB structure, while in panels (b) and (d) we visualize the hydrophobic ligand (trimethoprim for DHFR and myristic acid for LBD) in magenta. All images were generated using Pymol \cite{Pymol}.}
	\label{fig:drugA}
\end{figure}

\begin{figure}[!tb]
\centering
\includegraphics[width=0.9\textwidth]{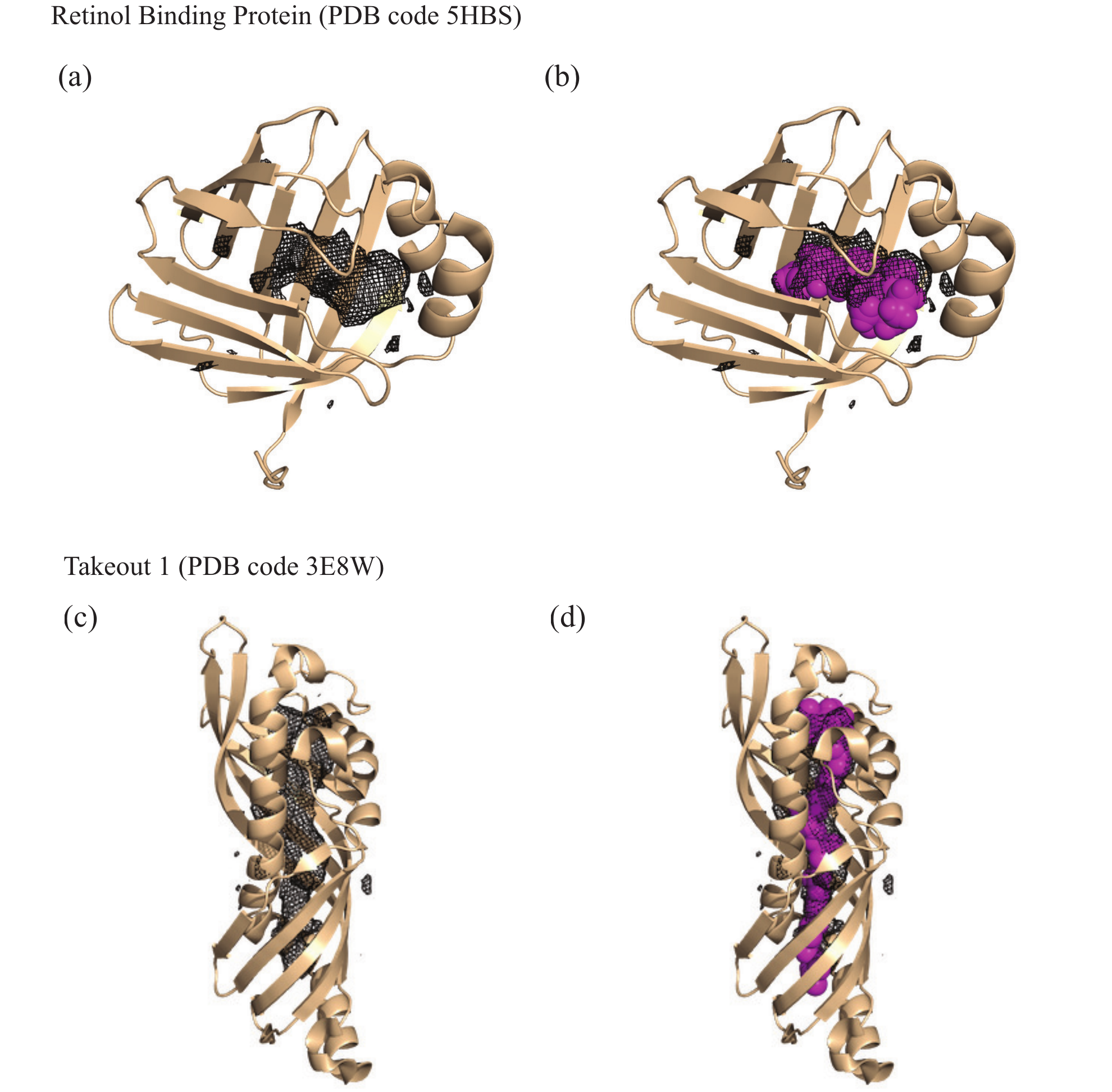}
	\caption{Hydrophobic occupancy maps, black mesh, superimposed onto the PDB structures for a retinol binding protein (RBP, PDB code 5HBS) and a TakeOut 1 protein (To1, PDB code 3E8W). The maps are derived from the
	densities of hydrophobic probes computed by AquaVit, and represented at +20 $\sigma$. Those maps are derived from the apo structure of the protein, i.e. in the absence of all crystallographic ligands and water molecules. In panels (a) and (c) we show the hydrophobic maps for RBP and To1 superposed to the apo PDB structure, while in panels (b) and (d) we visualize the hydrophobic ligand (trans retinol for RBP and 8-ubiquinone for To1) in magenta. All images were generated using Pymol \cite{Pymol}.}
	\label{fig:drugB}
\end{figure}

The set of proteins used for validation includes four proteins from different families and functions: a dihydrofolate reductase (DHFR) from Staphylococcus Aureus (PDB code 2W9H \cite{Heaslet:2009}),  a  maize lipid-transfer protein (PDB code 1FK2 \cite{Han:2001}), a human retinol binding protein 1 (PDB code 5HBS \cite{Silvaroli:2016}), and an insect Takeout 1 protein (PDB code 3E8W \cite{Hamiaux:2009}) .
The PDB structures correspond to those proteins bound to a hydrophobic ligand. All ligands, ions, and crystallographic water molecules, however, were removed prior to running AquaVit. In Figures \ref{fig:drugA} and \ref{fig:drugB}, we show the resulting hydrophobic probe occupancy maps superimposed on the PDB structures for all four proteins, with and without the ligand.

DHFR is the enzyme responsible for the NADPH‐dependent reduction of dihydrofolate to tetrahydrofolate, an essential cofactor in the synthesis of purines, methionine, and other key metabolites. Because of its importance in a wide range of cellular functions, DHFR has been the subject of much research targeting the enzyme for anticancer, antibacterial, and antimicrobial agents. Clinically used compounds targeting DHFR include methotrexate for the treatment of cancer and  trimethoprim (TMP) for the treatment of bacterial infections. The active site of DHFR is comprised of a large hydrophobic pocket which serves as the folate-binding site. This pocket was successfully detected by AquaVit, as illustrated in Figure \ref{fig:drugA}(a). This high density region of hydrophobic probes overlaps well with the TMP ligand that was co-crystallized with DHFR and bound within the hydrophobic active site \cite{Heaslet:2009} (see Figure \ref{fig:drugA}(b)).

Lipid binding proteins (LBP) facilitate the transfer of lipids between membrane. We consider a non specific LBP from maize, whose hydrophobic cavity can accommodate various lipids from C10 to C18. AquaVit was successful in identifying this cavity, as illustrated in Figure \ref{fig:drugA}(c). Interestingly, as we superimpose the ligand found in the PDB structure 1FK2, namely myristic acid, we find that the high density region of hydrophobic probes identified by AquaVit extends beyond this ligand (see Figure \ref{fig:drugA}(d)). Myristic acid is C14, while the pocket found by AquaVit indicates that the hydrophobic cavity of the maize LBP can accommodate larger ligand \cite{Han:2001}.

The cellular retinol-binding protein 1(CRBP1)  is another example of a protein with a large hydrophobic active site. CRBP1 is important in regulating
the uptake, storage and metabolism of retinoids (vitamin A and its derivatives). Its vitamin A binding site is lined with hydrophobic residues that are well conserved among retinol binding proteins. Those residues provide the non-polar interactions that stabilize the retinoid ligand. This pocket was successfully detected by AquaVit, as illustrated in Figure \ref{fig:drugB}(a). The high density region of hydrophobic probes overlaps well with the all trans retinol ligand that was co-crystallized with CRBP1 \cite{Silvaroli:2016}, as observed in Figure \ref{fig:drugB}(b).

Takeout (To) proteins are found exclusively in insects in which they have been proposed to play important roles in their physiology and behavior \cite{Saurabh:2018}.
Of particular interest to us, they have been suggested to bind to hydrophobic ligands. 
We considered the To 1 protein from \emph{Epiphyas postvittana}, a light brown apple moth, whose structure was solved by crystallography at 1.3 \AA \cite{Hamiaux:2009} in the presence of ubiquinone-8. The crystal structure revealed a 45 \AA long hydrophobic internal tunnel that extends to the full length of the protein. This pocket was successfully detected by AquaVit, as illustrated in Figure \ref{fig:drugB}(c). Note that the same pocket was originally characterized based
on geometry only \cite{Saurabh:2018}. AquaVit provides the additional information that this pocket is amenable to interaction with an hydrophobic ligand. This is confirmed as the high density region of hydrophobic probes overlaps well with the hydrophobic ubiquinone ligand, see Figure \ref{fig:drugB}(d).

\subsection{5.2 Environments of membrane proteins}

\begin{figure}[!tb]
\centering
\includegraphics[width=0.9\textwidth]{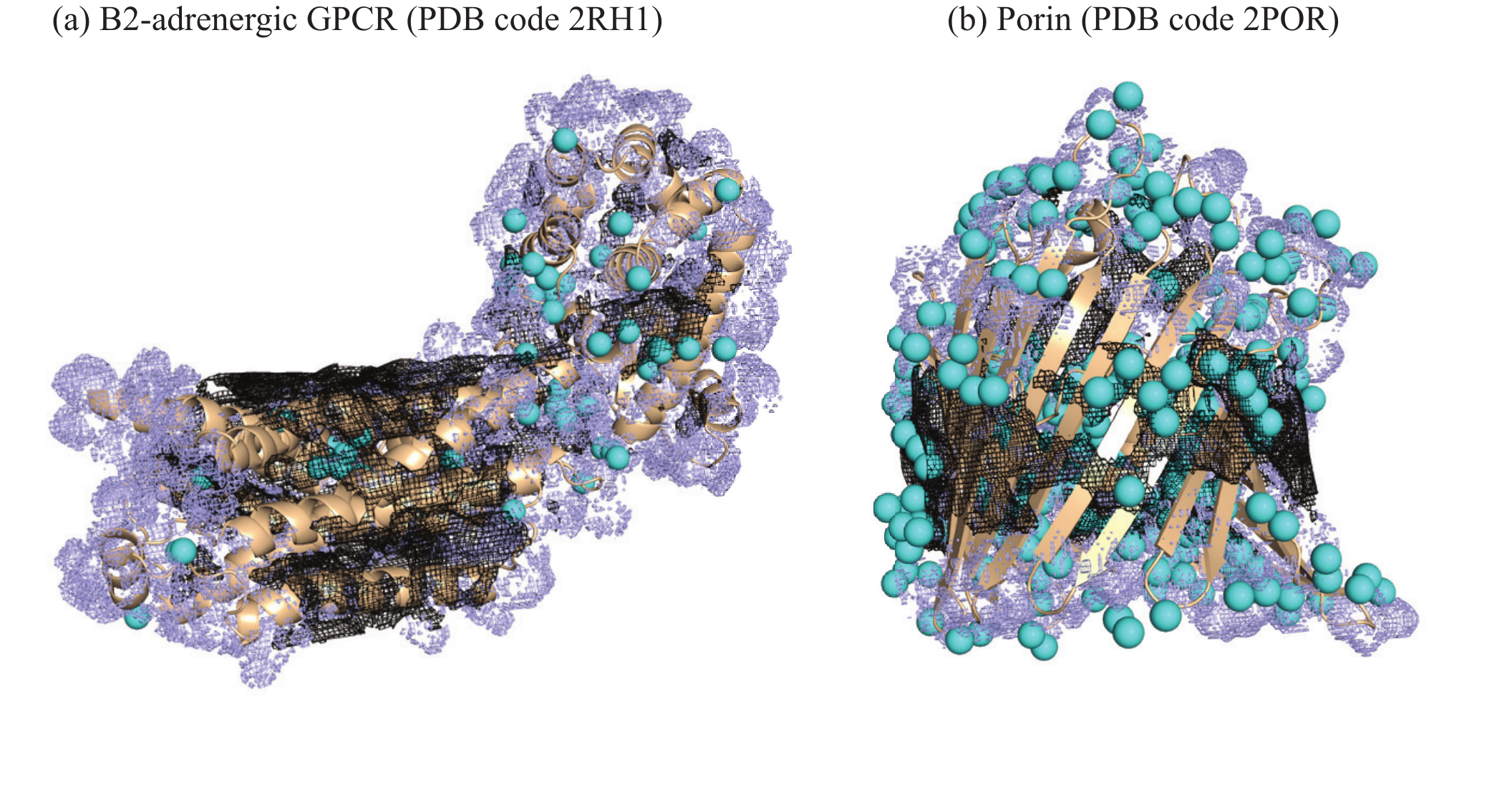}
	\caption{Hydrophobic occupancy maps, black mesh, and water dipole occupancy maps, blue mesh superimposed onto the PDB structures for (a) the $\beta_2$ adrenergic GPCR (PDB code 5HBS) and (b) a  porin (PDB code 2POR). The crystallographic waters are shown as cyan spheres. The maps are derived from the
	densities of hydrophobic probes and water dipoles computed by AquaVit, respectively, and represented at +10 $\sigma$ for the hydrophobic probes and at +0.3 $\sigma$ for the water dipoles. Those maps are derived from the apo structure of the proteins, i.e. in the absence of all crystallographic ligands and water molecules. }
	\label{fig:membrane}
\end{figure}

The subsection above illustrates that AquaVit is able to locate hydrophobic pockets in proteins. However, the solution of the HDPBL system of equations is more comprehensive and also provides information on dipolar density in the presence of a given salt concentration. Here we assess its ability to characterize the polar and non polar environments of membrane proteins. We consider two types of such protein, a member of the G protein coupled receptor (GPCR) family, with a transmembrane domain that consists of a helical bundle (PDB code 2RH1 \cite{Cherezov:2007}), and a porin that consist of a $\beta$-pleated sheet (PDB code 2POR \cite{Weiss:1992}). 
All ligands, ions, and crystallographic water molecules were removed prior to running AquaVit on those structures. In Figure \ref{fig:membrane} we show the resulting hydrophobic probe occupancy maps and water dipole occupancy maps, superimposed on the PDB structures, including the crystallographic water molecules.

$\beta_2$-adrenergic receptors ($\beta_2$AR) are members of the GPCR family that reside predominantly in smooth muscle. Their antagonists are used in particular in the treatment of asthma \cite{Bai:1992}. To study the structure of this membrane protein, Cherezov et al designed studied a chimera  consisting of $\beta_2$AR and the T4 Lysozyme (T4L) \cite{Cherezov:2007}. The crystal structure of this chimera reveals a fold for $\beta_2$AR composed of a transmembrane domain with a 7 helix bundle, and a standard all-helix fold for the T4L. Interactions between $\beta_2$AR an T4L are minimal. AquaVit provides a consistent image of the environment of this chimera, with a mostly hydrophobic environment for the transmembrane domain, and a mostly hydrophilic environment for T4L (see Figure \ref{fig:membrane} (a)). Interestingly, the water dipole occupancy map superimposes well with the crystallographic water moleculess detected in the structure.

Porins are integral membrane proteins that are found in the outer membrane of Gram-negative bacteria, mitochondria and chloroplasts. The crystal structure of the porin from \emph{Rhodobacter capsulatus} reveals a 16-stranded $\beta$-barrel, with all $\beta$-strands antiparallel and connected to their neighbors \cite{Weiss:1992}. AquaVit reveals a hydrophobic environment on the outside of the $\beta$-pleated sheet, with the loops between the strands in a more hydrophilic environment  (Figure \ref{fig:membrane} (b)). This is consistent with  the positions of the crystallographic water molecules that map with the water dipole density  map.

\subsection{5.3 Multiprobe analysis of a complex active site}

All examples presented above focused on the identification of hydrophobic pockets and characterization of the hydrophobic environment of proteins, based on the analyses of the hydrophobic occupancy maps generated by AquaVit. However, much akin to MixMD \cite{Graham:2018} and mLMMD \cite{Tan:2020}, AquaVit can be seen as a multiprobe analysis of the environment of
a protein. It generates the densities of water dipole, anion, cation, and hydrophobic molecules surrounding the protein that serve as probes to identify and characterize pockets in the protein. We use the ferredoxin oxidoreductase system to illustrate this functionality of AquaVit.

\begin{figure}[!tb]
\centering
\includegraphics[width=0.9\textwidth]{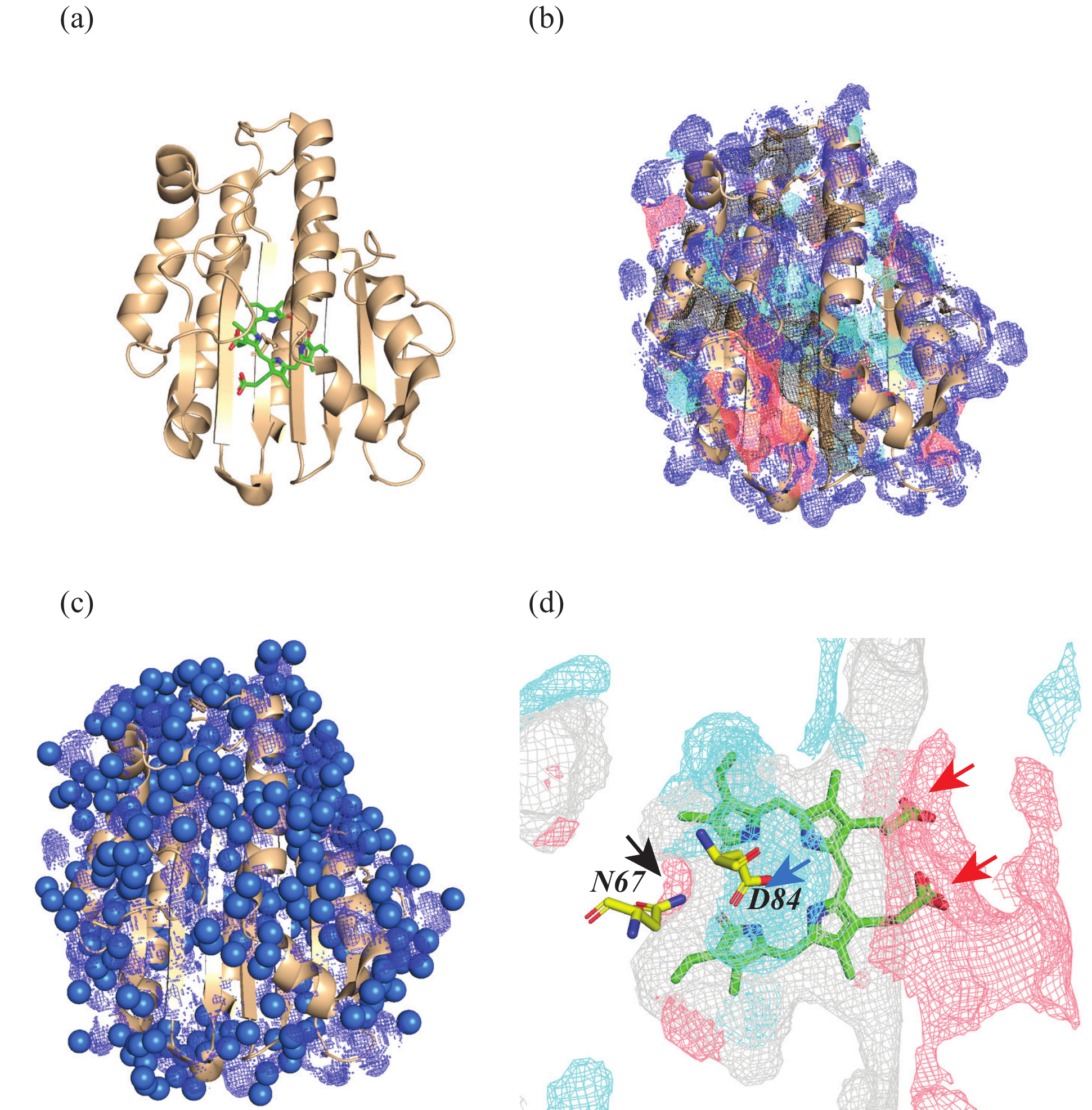}
	\caption{(a) Overall structure of PcbA, a ferredoxin-dependent bilin reductase (cartoon mode)  in presence of its substrate, biliverdin (BV): PDB code 1X9O \cite{Busch:2011}. (b) Superimposition of the occupancy maps of water (+0.3 $\sigma$, blue), hydrophobic probes (+10 $\sigma$, grey), anion (+20 $\sigma$, red) and cation (+20 $\sigma$, cyan) on the structure of PcbA, in the absence of BV.
	(c) The PcbA structure, the water occupancy map, and the crystallographic water molecules (shown as blue spheres).
	(d) Superposition of the occupancy maps of hydrophobic probes (grey), anion (red), and cation  (cyan) with the structure of BV (sticks). We also show the position of residues Asp84 and Asn67 that are known to play an important role in placing the ligand in its binding pocket \cite{Busch:2011}. Note the presence of anions density overlapping with the carboxyl groups of BV (red arrows) and forming a pocket in front of the N of the terminal group of Asn67. Note also the presence of cations around the carboxyl group of Asp84 (blue arrow).}
	\label{fig:pcy}
\end{figure}

Ferredoxin-dependent bilin reductases (FBDR) are enzymes that are involved in the reduction of biliverdin (BV) to form phycobilins used for light-perception or light harvesting in plants and cyanobacteria \cite{Dammeyer:2008}. 
Several members of this family have been identified in multiple species \cite{Frankenberg:2001}.
Among those, PebA reduces BV at its C15-C16 double bond to produce 15-16 dihydrobiliverdin (DHBV).
The structure of PebA from the cyanobacterium \emph{Synechococcus} sp. WH8020 with its substrate BV was determined at 1.55 \AA
(PDB code 1X9O \cite{Busch:2011}). This structure consists of a seven-stranded antiparallel $\beta$-sheet surrounded by six $\alpha$ helices (Figure \ref{fig:pcy}(a)). We used AquaVit to solve the HDPBL system of equations that characterize the environment of PeBA.
The calculation was performed on the protein structure alone, i.e. in the absence of the BV ligand and crystallographic water molecules.
Figure \ref{fig:pcy}(b) illustrates the densities of water dipoles, anions, cations, and hydrophobic probes around the structure of PebA in the form of occupancy maps. 
The whole protein is surrounded with water dipoles. This is in agreement with the fact that many crystallographic water molecules have been identified. Those molecules superimpose well with the water dipole occupancy map Figure \ref{fig:pcy}(c).
Of significant interest are the anion, cation, hydrophobic probe occupancy maps in the active site of PeBA; those
maps are illustrated in Figure \ref{fig:pcy}(d). The superposition of the ligand structure and of the two residues (Asp84 and Asn67) that define the
central polar centering ``pin" of the binding pocket \cite{Busch:2011} on those maps shows that the anion densities (in red)
map well with the two carboxyl groups on BV (red arrows). We also observe an anion pocket in front of the Nitrogen of the terminal group of Asn67 (black arrow). Note that the densities are expected to match with the chemical structure of the ligand, and be complementary to the chemical properties of the solute.
The hydrophobic probe density map superimposes well with the hydrophobic parts of BV.
Finally, we observe a cation occupancy in the region surrounding the 4 nitrogen of the rings of BV, as well as as a pocket in from of the carboxyl moiety of Asp84 (blue arrow).
These observations highlight the advantage of AquaVit to account for the multiple species that form the environment of a protein.

\subsection{5.4 Computing time}

\begin{figure}[!tb]
\centering
\includegraphics[width=0.6\textwidth]{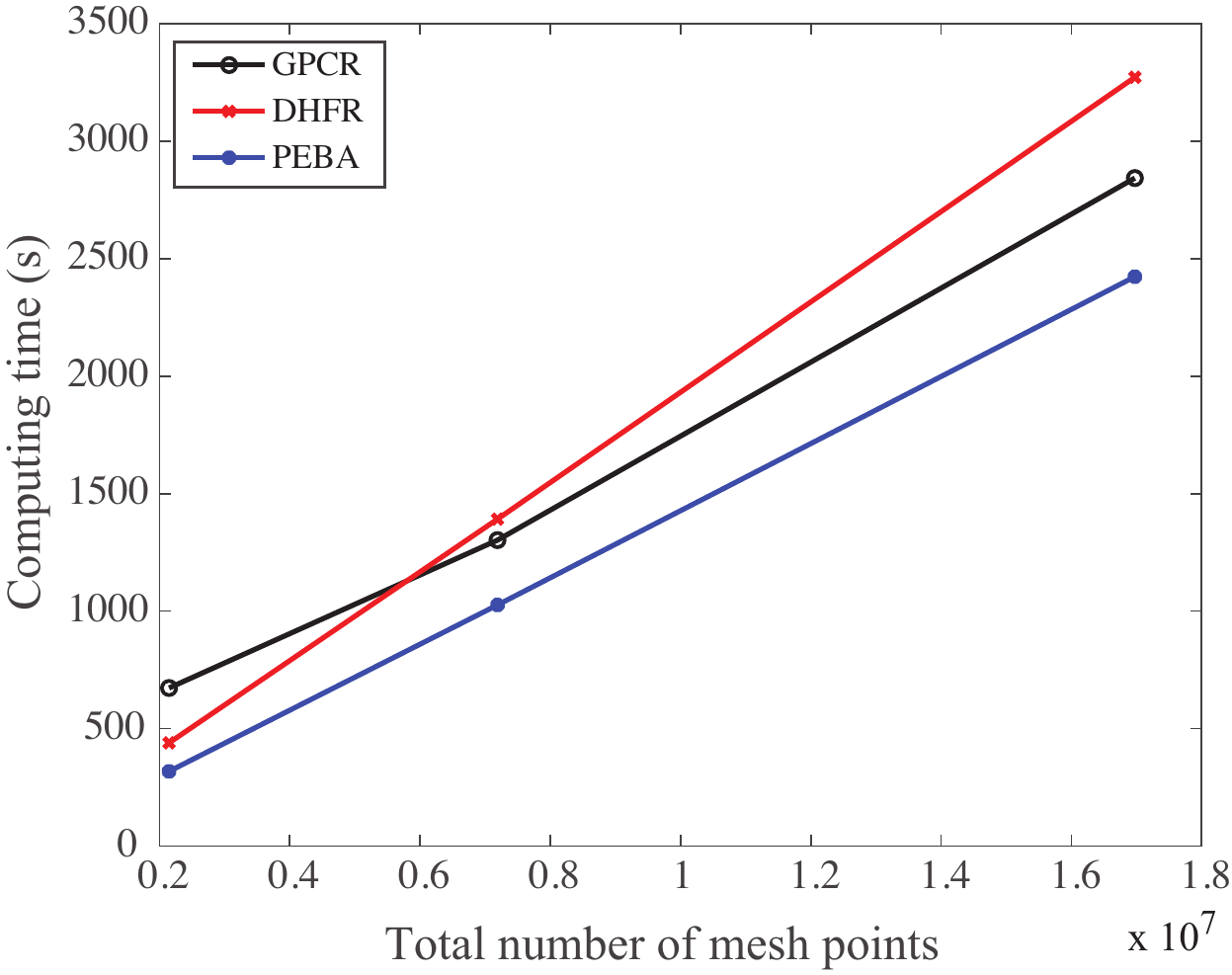}
	\caption{The computing time required to solve the system of equations HDPBL using AquaVit with a residual error lower than $TOL=10^{-4}$ (see text for details) is plotted versus the total number of points in the mesh for the three proteins DHFR (PDB code 2W9H), GPCR (PDB code 2RH1), and PEBA (PDB code 2X9O). 
	All computations were performed on an Intel Core i7 processor with 8 cores running at 4.00GHz, and 64GB of memory, although AquaVit was compiled without parallelization options.}
	\label{fig:cputime}
\end{figure}

In Figure \ref{fig:cputime} we report the computing times for solving the HDPBL system of equations using AquaVit for three systems, DHFR (PDB code 2W9H), GPCR (PDB code 2RH1), and PEBA (PDB code 2X9O) under the standard conditions defined in the Methods section. 
In addition to the inputs specific to the system under consideration (structure of the solute, concentrations of water dipoles, salt, and hydrophobic probes, dipole moment of the water dipole, and parameters of the Yukawa potential for hydrophobic interactions),
AquaVit relies on the parameters of algorithm \ref{alg:algorithm1} to solve the HDPBL system. These parameters include on the size of the Cartesian grid used to solve the PDEs, the tolerance TOL that serves as stopping criteria when solving the system self consistently, and the parameter $\lambda$ used to update the fields $u$ and $v$ (see Algorithm  \ref{alg:algorithm1} presented in section 3). In all calculations presented above, TOL is set to $10^{-4}$ and $\lambda$ is set to 0.7 (this number is somewhat arbitrary and could in fact be optimized for each system considered). We have also used consistently a grid of size $193^3$, but in figure \ref{fig:cputime} we report the computing time of AquaVit as a function of this grid size. 
We observe a near linear dependence of the computing times of AquaVit with respect to the total number of points in the grid, as expected \cite{Koehl:2010}. These computing time do not differ significantly between the three proteins, despite the fact that they are very different in size (157 residues for 2W9H, 237 residues for 2X9O, and 465 residues for 2RH1).  The average computing time for a grid of $193^3$ points is 1240 s, i.e. approximately 21 minutes. 
We note that the computing times reported are both CPU and clock time, i.e. AquaVit ran on a single core and does not benefit from parallelization.
We acknowledge that there is much room for improvement in the implementation of  algorithm \ref{alg:algorithm1} in AquaVit.

\subsection{5.5 Limitations of the HDBPL model: absence of flexibility}

We have shown above that the HDPBL model enables the discovery and characterization of binding sites on proteins. 
In addition, HDPBL is fast (see Figure \ref{fig:cputime}) or at least competitive with respect to computing time compared to the ligand mapping molecular dynamics simulations. It has one major limitation, however, as it considers the structure of the protein to be static. While it was
not an issue in the examples described in the previous sections, we illustrate here a case in which dynamics matter.

\begin{figure}[!tb]
\centering
\includegraphics[width=0.9\textwidth]{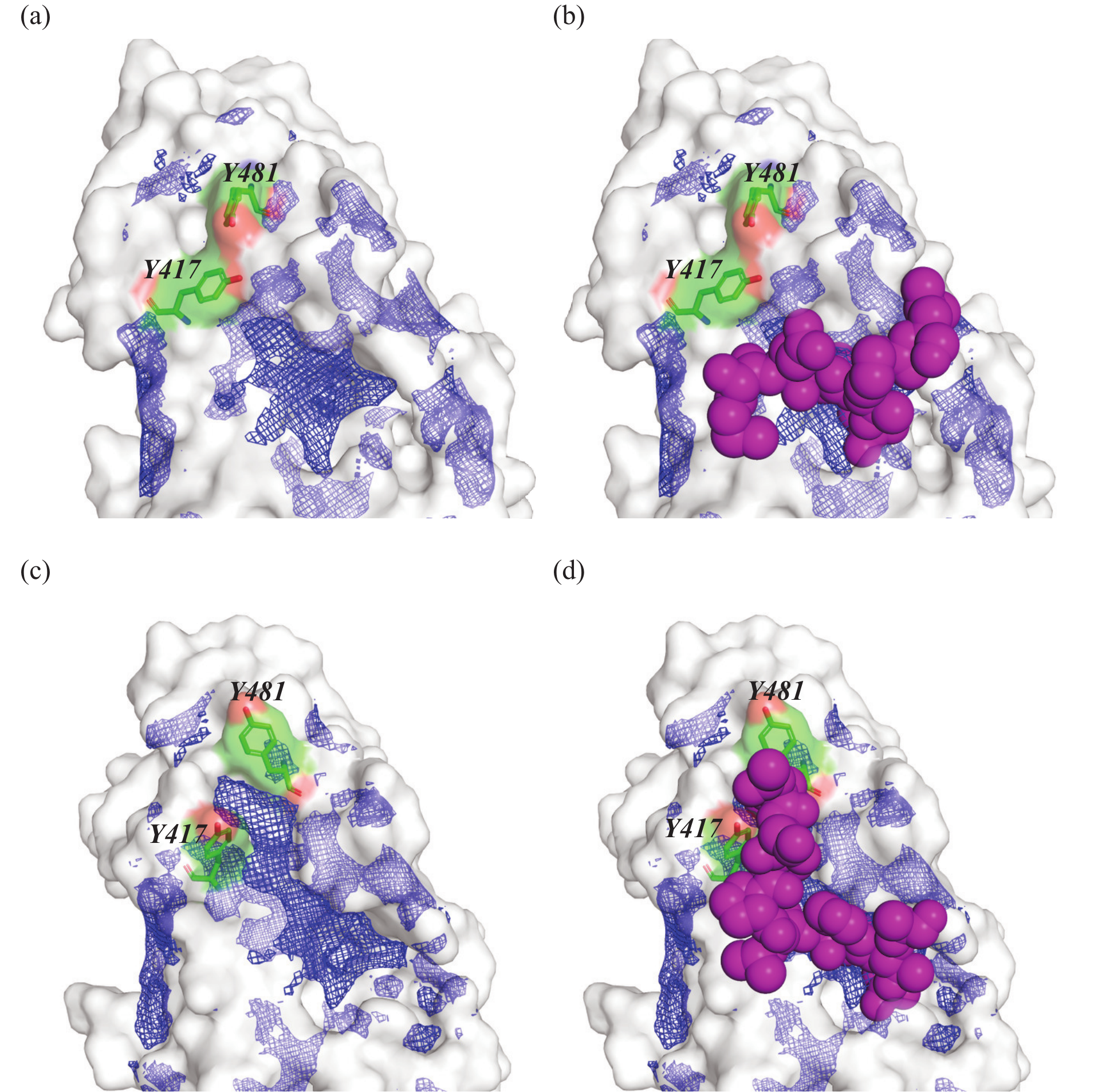}
	\caption{(a) Superimposition of the occupancy maps of  hydrophobic probes (+20 $\sigma$, blue) on the structure of the polo binding domain (PBD) of human polo like kinase 1 (PLK1) (PDB code 1Q4K). The occupancy map is computed using AquaVit, on the PBD structure in the absence of the ligand and crystallographic water molecules.
	(b) Same as in (a), but with the ligand (the phosphopeptide MetGlnSerpThrProLeu) shown in magenta. Note that the ligand
	overlaps well with one of the hydrophobic pockets identified by AquaVit.
	(c) Superimposition of the occupancy maps of  hydrophobic probes (+20 $\sigma$, blue) on the structure of the polo binding domain (PBD) of human polo like kinase 1 (PLK1) (PDB code 3P37K). The occupancy map is computed using AquaVit, on the PBD structure in the absence of the ligand and crystallographic water molecules. Note the longer hydrophobic pocket compared to 1Q4K, as a result
	of the change of conformation of Tyr417 and Tyr481.
	(d) Same as in (a), but with the ligand (the phosphopeptide PheAspProProLeuHisSerpThrAla) shown in magenta. The ligand
	overlaps well with the long hydrophobic pocket identified by AquaVit.
}
	\label{fig:pbd}
\end{figure}

Polo‐like kinases (PLKs) are a family of serine/threonine kinases related to the polo gene product of \emph{Drosophila melanogaster}.
Most PLKs have multiple functions which map with their organization in domains.
Their C-terminal regions, for example, contain the polo box domain (PBD), which  helps in their sub-cellular
localization by binding to serine- or threonine- phosphorylated sequences on target proteins. 
Cheng et al have determined the structure of the PBD domain of the human PLK1 in the presence of  a phosphopeptide
with sequence MetGlnSerpThrProLeu, where pThr indicates that the threonin is phosphorylated (PDB code 1Q4K \cite{Cheng:2003}). 
The phosphopeptide was found to bind on the surface of the protein, in an hydrophobic pocket.
We ran AquaVit on a single copy of the PBD domain in the absence of the peptide ligand and of all crystallographic water molecules.
We can identify the ligand binding hydrophobic pocket of PBD from the resulting hydrophobic probe density map, as illustrated in Figure \ref{fig:pbd}(a) and (b).

Interestingly, the binding of MetGlnSerpThrProLeu onto PBD was found to induce very little conformational changes \cite{Cheng:2003}.
Two subsequent independent studies \cite{Liu:2011, Sledz:2011}, however, identified a cryptic hydrophobic binding site that is close to the original phosphate binding site identified by Cheng et al \cite{Cheng:2003}.
The opening of this cryptic binding site is the result of a change in the conformation of the side chains of Tyr417 and Tyr481.
This opening enables PBD to bind longer phosphophopeptides.
Sledz et al \cite{Sledz:2011} for example presented the structure of the complex of PBD with the phosphopeptide PheAspProProLeuHisSerpThrAla (PDB code 3P37).
Interestingly, when we ran AquaVit on a single copy of the PBD domain in the absence of the peptide ligand and of all crystallographic waters in its configuration from 3P37, we were able to identify the longer ligand binding hydrophobic pocket of PBD which includes the cryptic pockets, which can bind this longer phosphopeptide. This is illustrated in Figure \ref{fig:pbd}(c) and (d).
AquaVit, however, was not able to detect this longer pocket from the structure in 1Q4K, as residues Tyr417 and Tyr481 were closing the cryptic pocket.

\section{6. Conclusion}
\label{sec:conclusion}

We have developed the HDBPL model and implemented it in the program AquaVit as a tool for identifying and characterizing binding sites in proteins. A protein of interest is immersed in a lattice gas containing water dipoles, anions, cations, and hydrophobic molecules. The charged molecules interact between themselves and with the solute charges through electrostatic interactions, while the hydrophobic molecules (including the non polar groups on the solute) interact based on a Yukawa potential. We impose steric constraints on the lattice, as well as incompressibility, i.e. all sites of the lattice are occupied. The system (protein of interest and its environment) is then characterized with an effective free energy that depends on two fields $\phi(\mathbf{r})$ and $\psi(\mathbf{r})$ corresponding to the electrostatics and hydrophobic interactions, respectively. The Euler-Lagrange equations obtained by minimizing the free energy with respect to those two fields form a system of two Poisson-Boltzmann like PDEs, HDPBL, which we solve using a self consistent approach, implemented in the program AquaVit. The outputs of HDPBL are the densities of the different species, and peaks of densities are expected to reveal the presence of compatible binding sites. We have tested and validated the ability of HDPBL to detect pockets that bind to hydrophobic ligands (the DHFR, a lipid binding protein, HIV protease, and a retinol binding protein), polar ligands (biliverdin), as well as to characterize the environment of membrane proteins such as GPCR and a porin. 

The HDPBL equations form a system of two coupled second order elliptic nonlinear PDE. While those equations are akin to the PB equation, they cannot be solved directly with a PB solver, mostly because the two equations are strongly dependent.
We have proposed, however, an algorithm that makes use of a standard PB solver by solving those equations self consistently. The same approach was used previously to solve the DBPL equation \cite{Koehl:2010}, as well as the YULP system of equation \cite{Koehl:2009}. This algorithm is relatively simple to implement and can be adapted to any PB solver, including those based on finite elements methods, which we did not consider here. We are currently developing a more versatile solver based on those methods.\\
In the current implementation of Aquavit, the lattice is populated with point-like electric charges, dipoles and hydrophobic molecules.
One could also include finite-size dipoles made of an electric and an hydrophobic charge, as well as an electric dipole attached to a hydrophobic charge.  In this case, these entities will react to an hydrophobic field $\nabla \Psi(\mathbf{r})$, using the same formalism. 
This could be well adapted to study the solvation of proteins in the presence of large and more complex co-solvents such as acetonitrile or DMSO.

AquaVit, our program for solving the HDPBL system of equations, is fast and as such compares favorably with the ligand mapping molecular dynamics simulations that have been designed with the same goal of detecting and characterizing binding sites in proteins.
The latter, however, have the significant advantage that they account for the dynamics of the solute. As such, they have been shown to detect cryptic bind sites in proteins \cite{Tan:2012, Kimura:2017, Schmidt:2019, Tan:2020}, namely sites that are not accessible unless a structural change occur in the protein.  In its current formulation, the HDPBL system of equations assumes that the protein is static; such conformational changes are then inaccessible. As an extension to HDBPL, one could model the dynamics of the solute protein with its low frequency normal modes, using for example a simple elastic model to compute those modes \cite{Tirion:1996, Sanejouand:2013, Voth:2013}. 
Ultimately, we want to develop AquaVit as a tool for structure-based as well as dynamics-based drug design.


\begin{acknowledgement}

The work discussed here originated from a visit by P.K. at the Institut de Physique Th\'{e}orique, CEA Saclay, France, 
during the fall of 2019. He thanks them for their hospitality and financial support. PK acknowledges support from the
University of California Multicampus Research Programs and Initiatives (grant No. M21PR3267).

\end{acknowledgement}




\begin{mcitethebibliography}{59}
\providecommand*\natexlab[1]{#1}
\providecommand*\mciteSetBstSublistMode[1]{}
\providecommand*\mciteSetBstMaxWidthForm[2]{}
\providecommand*\mciteBstWouldAddEndPuncttrue
  {\def\EndOfBibitem{\unskip.}}
\providecommand*\mciteBstWouldAddEndPunctfalse
  {\let\EndOfBibitem\relax}
\providecommand*\mciteSetBstMidEndSepPunct[3]{}
\providecommand*\mciteSetBstSublistLabelBeginEnd[3]{}
\providecommand*\EndOfBibitem{}
\mciteSetBstSublistMode{f}
\mciteSetBstMaxWidthForm{subitem}{(\alph{mcitesubitemcount})}
\mciteSetBstSublistLabelBeginEnd
  {\mcitemaxwidthsubitemform\space}
  {\relax}
  {\relax}

\bibitem[Shuker \latin{et~al.}(1996)Shuker, Hadjuk, Meadows, and
  Fesik]{Shuker:1996}
Shuker,~S.; Hadjuk,~P.; Meadows,~R.; Fesik,~S. Discovering high-affinity
  ligands for proteins: {SAR} by {NMR}. \emph{Science} \textbf{1996},
  \emph{274}, 1531--1534\relax
\mciteBstWouldAddEndPuncttrue
\mciteSetBstMidEndSepPunct{\mcitedefaultmidpunct}
{\mcitedefaultendpunct}{\mcitedefaultseppunct}\relax
\EndOfBibitem
\bibitem[Mattos and Ringe(1996)Mattos, and Ringe]{Mattos:1996}
Mattos,~C.; Ringe,~D. Locating and characterizing binding sites on proteins.
  \emph{Nat.\ Biotechnol.} \textbf{1996}, \emph{14}, 595--599\relax
\mciteBstWouldAddEndPuncttrue
\mciteSetBstMidEndSepPunct{\mcitedefaultmidpunct}
{\mcitedefaultendpunct}{\mcitedefaultseppunct}\relax
\EndOfBibitem
\bibitem[Hardy and Wells(2004)Hardy, and Wells]{Hardy:2004}
Hardy,~J.; Wells,~J. Searching for new allosteric sites in enzymes.
  \emph{Curr.\ Opin.\ Struct.\ Biol.} \textbf{2004}, \emph{14}, 706--715\relax
\mciteBstWouldAddEndPuncttrue
\mciteSetBstMidEndSepPunct{\mcitedefaultmidpunct}
{\mcitedefaultendpunct}{\mcitedefaultseppunct}\relax
\EndOfBibitem
\bibitem[Henrich \latin{et~al.}(2010)Henrich, Salo-Ahen, Huang, Rippmann,
  Cruciani, and Wade]{Henrich:2010}
Henrich,~S.; Salo-Ahen,~O.; Huang,~B.; Rippmann,~F.; Cruciani,~G.; Wade,~R.
  Computational approaches to identifying and characterizing protein binding
  sites for ligand design. \emph{J.\ Molec.\ Recognit.} \textbf{2010},
  \emph{23}, 209--219\relax
\mciteBstWouldAddEndPuncttrue
\mciteSetBstMidEndSepPunct{\mcitedefaultmidpunct}
{\mcitedefaultendpunct}{\mcitedefaultseppunct}\relax
\EndOfBibitem
\bibitem[Konc and Jane\v{z}i\v{c}(2014)Konc, and Jane\v{z}i\v{c}]{Konc:2014}
Konc,~J.; Jane\v{z}i\v{c},~D. Binding site comparison for function prediction
  and pharmaceutical discovery. \emph{Curr.\ Opin.\ Struct.\ Biol.}
  \textbf{2014}, \emph{25}, 34--39\relax
\mciteBstWouldAddEndPuncttrue
\mciteSetBstMidEndSepPunct{\mcitedefaultmidpunct}
{\mcitedefaultendpunct}{\mcitedefaultseppunct}\relax
\EndOfBibitem
\bibitem[Zhao \latin{et~al.}(2020)Zhao, Cao, and Zhang]{Zhao:2020}
Zhao,~J.; Cao,~Y.; Zhang,~L. Exploring the computational methods for
  protein-ligand binding site prediction. \emph{Comput.\ Struct.\ Biotechnol.\
  J.} \textbf{2020}, \emph{18}, 417--426\relax
\mciteBstWouldAddEndPuncttrue
\mciteSetBstMidEndSepPunct{\mcitedefaultmidpunct}
{\mcitedefaultendpunct}{\mcitedefaultseppunct}\relax
\EndOfBibitem
\bibitem[Ivetac and McCammon(2018)Ivetac, and McCammon]{Iveta:2012}
Ivetac,~A.; McCammon,~J. A molecular dynamics ensemble-based approach for the
  mapping of druggable binding sites. \emph{Methods Mol.\ Biol.} \textbf{2018},
  \emph{819}, 3--12\relax
\mciteBstWouldAddEndPuncttrue
\mciteSetBstMidEndSepPunct{\mcitedefaultmidpunct}
{\mcitedefaultendpunct}{\mcitedefaultseppunct}\relax
\EndOfBibitem
\bibitem[Feng and Barakat(2018)Feng, and Barakat]{Feng:2018}
Feng,~T.; Barakat,~K. Molecular dynamics simulation and prediction of druggable
  binding sites. \emph{Methods Mol.\ Biol.} \textbf{2018}, \emph{1762},
  87--103\relax
\mciteBstWouldAddEndPuncttrue
\mciteSetBstMidEndSepPunct{\mcitedefaultmidpunct}
{\mcitedefaultendpunct}{\mcitedefaultseppunct}\relax
\EndOfBibitem
\bibitem[\'{S}led\'{z} and Caflisch(2018)\'{S}led\'{z}, and
  Caflisch]{Sledz:2018}
\'{S}led\'{z},~P.; Caflisch,~A. Protein structure-based drug design: from
  docking to molecular dynamics. \emph{Curr.\ Opin.\ Struct.\ Biol.}
  \textbf{2018}, \emph{48}, 93--102\relax
\mciteBstWouldAddEndPuncttrue
\mciteSetBstMidEndSepPunct{\mcitedefaultmidpunct}
{\mcitedefaultendpunct}{\mcitedefaultseppunct}\relax
\EndOfBibitem
\bibitem[Basse \latin{et~al.}(2010)Basse, Kaar, Settanni, Joerger, Rutherford,
  and Fersht]{Basse:2010}
Basse,~N.; Kaar,~J.; Settanni,~G.; Joerger,~A.; Rutherford,~T.; Fersht,~A.
  Toward the rational design of p53-stabilizing drugs: Probing the surface of
  the oncogenic {Y220C} mutant. \emph{Chem.\ Biol.} \textbf{2010}, \emph{17},
  46--56\relax
\mciteBstWouldAddEndPuncttrue
\mciteSetBstMidEndSepPunct{\mcitedefaultmidpunct}
{\mcitedefaultendpunct}{\mcitedefaultseppunct}\relax
\EndOfBibitem
\bibitem[Tan \latin{et~al.}(2012)Tan, Sledz, Lang, Stubbs, Spring, Abell, and
  Best]{Tan:2012}
Tan,~Y.; Sledz,~P.; Lang,~S.; Stubbs,~C.; Spring,~D.; Abell,~C.; Best,~R. Using
  ligand-mapping simulations to design a ligand selectively targeting a cryptic
  surface pocket of {P}olo-Like kinase 1. \emph{Angew.\ Chem.,\ Int.\ Ed.}
  \textbf{2012}, \emph{51}, 10078--10081\relax
\mciteBstWouldAddEndPuncttrue
\mciteSetBstMidEndSepPunct{\mcitedefaultmidpunct}
{\mcitedefaultendpunct}{\mcitedefaultseppunct}\relax
\EndOfBibitem
\bibitem[Tan \latin{et~al.}(2014)Tan, Spring, and Verma]{Tan:2014}
Tan,~Y.; Spring,~D.; Verma,~C. The use of chlorobenzene as a probe molecule in
  molecular dynamics simulations. \emph{J.\ Chem.\ Inf.\ Model.} \textbf{2014},
  \emph{54}, 1821--1827\relax
\mciteBstWouldAddEndPuncttrue
\mciteSetBstMidEndSepPunct{\mcitedefaultmidpunct}
{\mcitedefaultendpunct}{\mcitedefaultseppunct}\relax
\EndOfBibitem
\bibitem[Kalenkiewicz \latin{et~al.}(2015)Kalenkiewicz, Grant, and
  Yang]{Kalenkiewicz:2015}
Kalenkiewicz,~A.; Grant,~B.; Yang,~C.-Y. Enrichment of druggable conformations
  from apo protein structures using cosolvent-accelerated molecular dynamics.
  \emph{Biology} \textbf{2015}, \emph{4}, 344--366\relax
\mciteBstWouldAddEndPuncttrue
\mciteSetBstMidEndSepPunct{\mcitedefaultmidpunct}
{\mcitedefaultendpunct}{\mcitedefaultseppunct}\relax
\EndOfBibitem
\bibitem[Kimura \latin{et~al.}(2017)Kimura, Hu, Ruvinsky, Sherman, and
  Favia]{Kimura:2017}
Kimura,~S.; Hu,~H.; Ruvinsky,~A.; Sherman,~W.; Favia,~A. Deciphering cryptic
  binding sites on proteins by mixed-solvent molecular dynamics. \emph{J.\
  Chem.\ Inf.\ Model.} \textbf{2017}, \emph{57}, 1388--1401\relax
\mciteBstWouldAddEndPuncttrue
\mciteSetBstMidEndSepPunct{\mcitedefaultmidpunct}
{\mcitedefaultendpunct}{\mcitedefaultseppunct}\relax
\EndOfBibitem
\bibitem[Schmidt \latin{et~al.}(2019)Schmidt, Boehm, McClendon, Torella, and
  Gohlke]{Schmidt:2019}
Schmidt,~D.; Boehm,~M.; McClendon,~C.; Torella,~R.; Gohlke,~H.
  Cosolvent-enhanced sampling and unbiased identification of cryptic pockets
  suitable for structure-based drug design. \emph{J.\ Chem.\ Theory Comput.}
  \textbf{2019}, \emph{15}, 3331--3343\relax
\mciteBstWouldAddEndPuncttrue
\mciteSetBstMidEndSepPunct{\mcitedefaultmidpunct}
{\mcitedefaultendpunct}{\mcitedefaultseppunct}\relax
\EndOfBibitem
\bibitem[Nerenberg and Head-Gordon(2018)Nerenberg, and
  Head-Gordon]{Nerenberg:2018}
Nerenberg,~P.; Head-Gordon,~T. New developments in force fields for
  biomolecular simulations. \emph{Curr.\ Opin.\ Struct.\ Biol.} \textbf{2018},
  \emph{49}, 129--138\relax
\mciteBstWouldAddEndPuncttrue
\mciteSetBstMidEndSepPunct{\mcitedefaultmidpunct}
{\mcitedefaultendpunct}{\mcitedefaultseppunct}\relax
\EndOfBibitem
\bibitem[Huang and Jr.(2018)Huang, and Jr.]{Huang:2018}
Huang,~J.; Jr.,~A. D.~M. Force field development and simulations of
  intrinsically disordered proteins. \emph{Curr.\ Opin.\ Struct.\ Biol.}
  \textbf{2018}, \emph{48}, 40--48\relax
\mciteBstWouldAddEndPuncttrue
\mciteSetBstMidEndSepPunct{\mcitedefaultmidpunct}
{\mcitedefaultendpunct}{\mcitedefaultseppunct}\relax
\EndOfBibitem
\bibitem[Inakollu \latin{et~al.}(2020)Inakollu, Geerke, Rowley, and
  H]{Inakollu:2020}
Inakollu,~V.; Geerke,~D.; Rowley,~C.; H,~H.~Y. Polarisable force fields: what
  do they add in biomolecular simulations? \emph{Curr.\ Opin.\ Struct.\ Biol.}
  \textbf{2020}, \emph{61}, 182--190\relax
\mciteBstWouldAddEndPuncttrue
\mciteSetBstMidEndSepPunct{\mcitedefaultmidpunct}
{\mcitedefaultendpunct}{\mcitedefaultseppunct}\relax
\EndOfBibitem
\bibitem[{van der Spoel}(2020)]{Spoel:2020}
{van der Spoel},~D. Systematic design of biomolecular force fields.
  \emph{Curr.\ Opin.\ Struct.\ Biol.} \textbf{2020}, \emph{67}, 18--24\relax
\mciteBstWouldAddEndPuncttrue
\mciteSetBstMidEndSepPunct{\mcitedefaultmidpunct}
{\mcitedefaultendpunct}{\mcitedefaultseppunct}\relax
\EndOfBibitem
\bibitem[Grochowski and Trylska(2007)Grochowski, and Trylska]{Grochowski:2007}
Grochowski,~P.; Trylska,~J. Continuum molecular electrostatics, salt effects,
  and counterion binding - {A} review of the {Poisson-Boltzmann} theory and its
  modifications. \emph{Biopolymers} \textbf{2007}, \emph{89}, 93--113\relax
\mciteBstWouldAddEndPuncttrue
\mciteSetBstMidEndSepPunct{\mcitedefaultmidpunct}
{\mcitedefaultendpunct}{\mcitedefaultseppunct}\relax
\EndOfBibitem
\bibitem[Borukhov \latin{et~al.}(1997)Borukhov, Andelman, and
  Orland]{Borukhov:1997}
Borukhov,~I.; Andelman,~D.; Orland,~H. Steric effects in electrolytes: A
  modified {Poisson-Boltzmann} equation. \emph{Phys.\ Rev.\ Lett.}
  \textbf{1997}, \emph{79}, 435--438\relax
\mciteBstWouldAddEndPuncttrue
\mciteSetBstMidEndSepPunct{\mcitedefaultmidpunct}
{\mcitedefaultendpunct}{\mcitedefaultseppunct}\relax
\EndOfBibitem
\bibitem[Chu \latin{et~al.}(2007)Chu, Bai, Lipfert, Herschlag, and
  Doniach]{Chu:2007}
Chu,~V.; Bai,~Y.; Lipfert,~J.; Herschlag,~D.; Doniach,~S. Evaluation of ion
  binding to {DNA} duplexes using a size-modified {Poisson-Boltzmann} theory.
  \emph{Biophys. J.} \textbf{2007}, \emph{93}, 3202--3209\relax
\mciteBstWouldAddEndPuncttrue
\mciteSetBstMidEndSepPunct{\mcitedefaultmidpunct}
{\mcitedefaultendpunct}{\mcitedefaultseppunct}\relax
\EndOfBibitem
\bibitem[Xie \latin{et~al.}(2020)Xie, Audi, and Dash]{Xie:2020}
Xie,~D.; Audi,~S.; Dash,~R. A size modified {P}oisson-{B}oltzmann ion channel
  model in a solvent of multiple ionic species: application to
  voltage-dependent anion channel. \emph{J.\ Comp.\ Chem.} \textbf{2020},
  \emph{41}, 218--230\relax
\mciteBstWouldAddEndPuncttrue
\mciteSetBstMidEndSepPunct{\mcitedefaultmidpunct}
{\mcitedefaultendpunct}{\mcitedefaultseppunct}\relax
\EndOfBibitem
\bibitem[Stein \latin{et~al.}(2019)Stein, Herbert, and Head-Gordon]{Stein:2019}
Stein,~C.; Herbert,~J.; Head-Gordon,~M. The {P}oisson-{B}oltzmann model for
  implicit solvation of electrolyte solutions: Quantum chemical implementation
  and assessment via Sechenov coefficients. \emph{J.\ Chem.\ Phys.}
  \textbf{2019}, \emph{151}, 224111\relax
\mciteBstWouldAddEndPuncttrue
\mciteSetBstMidEndSepPunct{\mcitedefaultmidpunct}
{\mcitedefaultendpunct}{\mcitedefaultseppunct}\relax
\EndOfBibitem
\bibitem[Pang and Zhou(2013)Pang, and Zhou]{Pang:2013}
Pang,~X.; Zhou,~H.-X. {P}oisson-{B}oltzmann calculations: van der Waals or
  molecular surface? \emph{Commun.\ Comput.\ Phys.} \textbf{2013}, \emph{13},
  1--12\relax
\mciteBstWouldAddEndPuncttrue
\mciteSetBstMidEndSepPunct{\mcitedefaultmidpunct}
{\mcitedefaultendpunct}{\mcitedefaultseppunct}\relax
\EndOfBibitem
\bibitem[Grant \latin{et~al.}(2001)Grant, Pickup, and Nicholls]{Grant:2001}
Grant,~J.; Pickup,~B.; Nicholls,~A. A smooth permittivity function for
  Poisson-Boltzmann solvation methods. \emph{J.\ Comp.\ Chem.} \textbf{2001},
  \emph{22}, 608--640\relax
\mciteBstWouldAddEndPuncttrue
\mciteSetBstMidEndSepPunct{\mcitedefaultmidpunct}
{\mcitedefaultendpunct}{\mcitedefaultseppunct}\relax
\EndOfBibitem
\bibitem[Azuara \latin{et~al.}(2006)Azuara, Lindahl, Koehl, Orland, and
  Delarue]{Azuara:2006}
Azuara,~C.; Lindahl,~E.; Koehl,~P.; Orland,~H.; Delarue,~M. Incorporating
  dipolar solvents with variable density in the {Poisson-Boltzmann} treatment
  of macromolecule electrostatics. \emph{Nucleic Acids Res.} \textbf{2006},
  \emph{34}, W34--W42\relax
\mciteBstWouldAddEndPuncttrue
\mciteSetBstMidEndSepPunct{\mcitedefaultmidpunct}
{\mcitedefaultendpunct}{\mcitedefaultseppunct}\relax
\EndOfBibitem
\bibitem[Abrashkin \latin{et~al.}(2007)Abrashkin, Andelman, and
  Orland]{Abrashkin:2007}
Abrashkin,~A.; Andelman,~D.; Orland,~H. Dipolar {Poisson-Boltzmann} equation:
  ions and dipoles close to charge interfaces. \emph{Phys.\ Rev.\ Lett.}
  \textbf{2007}, \emph{99}, 77801\relax
\mciteBstWouldAddEndPuncttrue
\mciteSetBstMidEndSepPunct{\mcitedefaultmidpunct}
{\mcitedefaultendpunct}{\mcitedefaultseppunct}\relax
\EndOfBibitem
\bibitem[Azuara \latin{et~al.}(2008)Azuara, Orland, Bon, Koehl, and
  Delarue]{Azuara:2008}
Azuara,~C.; Orland,~H.; Bon,~M.; Koehl,~P.; Delarue,~M. Incorporating dipolar
  solvents with variable density in Poisson-Boltzmann electrostatics.
  \emph{Biophys.\ J.} \textbf{2008}, \emph{95}, 5587--5605\relax
\mciteBstWouldAddEndPuncttrue
\mciteSetBstMidEndSepPunct{\mcitedefaultmidpunct}
{\mcitedefaultendpunct}{\mcitedefaultseppunct}\relax
\EndOfBibitem
\bibitem[Koehl \latin{et~al.}(2009)Koehl, Orland, and Delarue]{Koehl:2009}
Koehl,~P.; Orland,~H.; Delarue,~M. Beyond {Poisson-Boltzmann}: modeling
  biomolecule-water and water-water interactions. \emph{Phys.\ Rev.\ Lett.}
  \textbf{2009}, \emph{102}, 087801\relax
\mciteBstWouldAddEndPuncttrue
\mciteSetBstMidEndSepPunct{\mcitedefaultmidpunct}
{\mcitedefaultendpunct}{\mcitedefaultseppunct}\relax
\EndOfBibitem
\bibitem[Koehl and Delarue(2010)Koehl, and Delarue]{Koehl:2010}
Koehl,~P.; Delarue,~M. Aquasol: an efficient solver for the dipolar
  {P}oisson-{B}oltzmann-{L}angevin equation. \emph{J.\ Chem.\ Phys.}
  \textbf{2010}, \emph{132}, 064101\relax
\mciteBstWouldAddEndPuncttrue
\mciteSetBstMidEndSepPunct{\mcitedefaultmidpunct}
{\mcitedefaultendpunct}{\mcitedefaultseppunct}\relax
\EndOfBibitem
\bibitem[Holst(1993)]{Holst_thesis}
Holst,~M. Multilevel Methods for the {Poisson-Boltzmann} Equation. Ph.D.\
  thesis, University of Illinois at Urbana-Champaign, USA, 1993\relax
\mciteBstWouldAddEndPuncttrue
\mciteSetBstMidEndSepPunct{\mcitedefaultmidpunct}
{\mcitedefaultendpunct}{\mcitedefaultseppunct}\relax
\EndOfBibitem
\bibitem[Holst and Saied(1995)Holst, and Saied]{Holst:1995}
Holst,~M.; Saied,~F. Numerical solution of the nonlinear {Poisson-Boltzmann}
  equation: developing more robust and efficient methods. \emph{J.\ Comp.\
  Chem.} \textbf{1995}, \emph{16}, 337--364\relax
\mciteBstWouldAddEndPuncttrue
\mciteSetBstMidEndSepPunct{\mcitedefaultmidpunct}
{\mcitedefaultendpunct}{\mcitedefaultseppunct}\relax
\EndOfBibitem
\bibitem[Berman \latin{et~al.}(2000)Berman, Westbrook, Feng, Gilliland, Bhat,
  Weissig, Shindyalov, and Bourne]{Berstein:2000}
Berman,~H.~M.; Westbrook,~J.; Feng,~Z.; Gilliland,~G.; Bhat,~T.~N.;
  Weissig,~H.; Shindyalov,~I.~N.; Bourne,~P.~E. The {P}rotein {D}ata {B}ank.
  \emph{Nucl.\ Acids.\ Res.} \textbf{2000}, \emph{28}, 235--242\relax
\mciteBstWouldAddEndPuncttrue
\mciteSetBstMidEndSepPunct{\mcitedefaultmidpunct}
{\mcitedefaultendpunct}{\mcitedefaultseppunct}\relax
\EndOfBibitem
\bibitem[Dolinsky \latin{et~al.}(2004)Dolinsky, Nielsen, Cammon, and
  Baker]{Dolinsky:2004}
Dolinsky,~T.; Nielsen,~J.; Cammon,~J.~M.; Baker,~N. {PDB2PQR}: an automated
  pipeline for the setup of {Poisson-Boltzmann} electrostatics calculations.
  \emph{Nucl.\ Acids.\ Res.} \textbf{2004}, \emph{32}, W665--W667\relax
\mciteBstWouldAddEndPuncttrue
\mciteSetBstMidEndSepPunct{\mcitedefaultmidpunct}
{\mcitedefaultendpunct}{\mcitedefaultseppunct}\relax
\EndOfBibitem
\bibitem[Sitkoff \latin{et~al.}(1994)Sitkoff, Sharp, and Honig]{Sitkoff:1994}
Sitkoff,~D.; Sharp,~K.; Honig,~B. Accurate calculation of hydration free
  energies using macroscopic solvent models. \emph{J.\ Phys.\ Chem.}
  \textbf{1994}, \emph{98}, 1978--1988\relax
\mciteBstWouldAddEndPuncttrue
\mciteSetBstMidEndSepPunct{\mcitedefaultmidpunct}
{\mcitedefaultendpunct}{\mcitedefaultseppunct}\relax
\EndOfBibitem
\bibitem[Guvench and {MacKerell Jr}(2009)Guvench, and {MacKerell
  Jr}]{Guvench:2009}
Guvench,~O.; {MacKerell Jr},~A. Computational fragment-based binding site
  identification by ligand competitive saturation. \emph{{PLoS} Comput.\ Biol.}
  \textbf{2009}, \emph{5}, e1000435\relax
\mciteBstWouldAddEndPuncttrue
\mciteSetBstMidEndSepPunct{\mcitedefaultmidpunct}
{\mcitedefaultendpunct}{\mcitedefaultseppunct}\relax
\EndOfBibitem
\bibitem[Oleinikovas \latin{et~al.}(2016)Oleinikovas, Saladino, Cossins, and
  Gervasio]{Oleinikovas:2016}
Oleinikovas,~V.; Saladino,~G.; Cossins,~B.; Gervasio,~F. Understanding cryptic
  pocket formation in protein targets by enhanced sampling simulations.
  \emph{J.\ Am.\ Chem.\ Soc.} \textbf{2016}, \emph{138}, 14257--14263\relax
\mciteBstWouldAddEndPuncttrue
\mciteSetBstMidEndSepPunct{\mcitedefaultmidpunct}
{\mcitedefaultendpunct}{\mcitedefaultseppunct}\relax
\EndOfBibitem
\bibitem[Heaslet \latin{et~al.}(2009)Heaslet, Harris, Fahnoe, Sarver, Putz,
  Chang, Subramanian, Barreiro, and Miller]{Heaslet:2009}
Heaslet,~H.; Harris,~M.; Fahnoe,~K.; Sarver,~R.; Putz,~H.; Chang,~J.;
  Subramanian,~C.; Barreiro,~G.; Miller,~J. Structural comparison of
  chromosomal and exogenous dihydrofolate reductase from \emph{Staphylococcus
  aureus} in complex with the potent inhibitor trimethoprim. \emph{Proteins:
  Struct.\ Func.\ Bioinfo.} \textbf{2009}, \emph{76}, 706--717\relax
\mciteBstWouldAddEndPuncttrue
\mciteSetBstMidEndSepPunct{\mcitedefaultmidpunct}
{\mcitedefaultendpunct}{\mcitedefaultseppunct}\relax
\EndOfBibitem
\bibitem[Han \latin{et~al.}(2001)Han, Lee, Song, Chang, Min, Moon, Shin, Kopka,
  Sawaya, Yuan, Kim, Choe, Lim, Moon, and Suh]{Han:2001}
Han,~G.; Lee,~J.; Song,~H.; Chang,~C.; Min,~K.; Moon,~J.; Shin,~D.; Kopka,~M.;
  Sawaya,~M.; Yuan,~H. \latin{et~al.}  Structural basis of non-specific lipid
  binding in maize lipid transfer protein complexes revealed by high-resolution
  X-ray crystallography. \emph{J.\ Mol.\ Biol.} \textbf{2001}, \emph{308},
  263--278\relax
\mciteBstWouldAddEndPuncttrue
\mciteSetBstMidEndSepPunct{\mcitedefaultmidpunct}
{\mcitedefaultendpunct}{\mcitedefaultseppunct}\relax
\EndOfBibitem
\bibitem[Silvaroli \latin{et~al.}(2016)Silvaroli, Arne, Chelstowska, Kiser,
  Banerjee, and Golczak]{Silvaroli:2016}
Silvaroli,~J.; Arne,~J.; Chelstowska,~S.; Kiser,~P.; Banerjee,~S.; Golczak,~M.
  Ligand binding induces conformational changes in human cellular
  retinol-binding protein 1 ({CRBP1}) revealed by atomic resolution crystal
  structures. \emph{J.\ Biol.\ Chem.} \textbf{2016}, \emph{291},
  8528--8540\relax
\mciteBstWouldAddEndPuncttrue
\mciteSetBstMidEndSepPunct{\mcitedefaultmidpunct}
{\mcitedefaultendpunct}{\mcitedefaultseppunct}\relax
\EndOfBibitem
\bibitem[Hamiaux \latin{et~al.}(2009)Hamiaux, Stanley, Greenwood, Baker, and
  Newcomb]{Hamiaux:2009}
Hamiaux,~C.; Stanley,~D.; Greenwood,~D.; Baker,~E.; Newcomb,~R.~D. Crystal
  structure of \emph{{E}piphyas postvittana} takeout 1 with bound ubiquinone
  supports a role as ligand carriers for takeout proteins in insects. \emph{J.\
  Biol.\ Chem.} \textbf{2009}, \emph{284}, 3496--3503\relax
\mciteBstWouldAddEndPuncttrue
\mciteSetBstMidEndSepPunct{\mcitedefaultmidpunct}
{\mcitedefaultendpunct}{\mcitedefaultseppunct}\relax
\EndOfBibitem
\bibitem[{Schr\"odinger, LLC}(2015)]{Pymol}
{Schr\"odinger, LLC}, The {PyMOL} Molecular Graphics System, Version~1.8.
  2015\relax
\mciteBstWouldAddEndPuncttrue
\mciteSetBstMidEndSepPunct{\mcitedefaultmidpunct}
{\mcitedefaultendpunct}{\mcitedefaultseppunct}\relax
\EndOfBibitem
\bibitem[Saurabh \latin{et~al.}(2018)Saurabh, Vanaphan, Wen, and
  Dauwalder]{Saurabh:2018}
Saurabh,~S.; Vanaphan,~N.; Wen,~W.; Dauwalder,~B. High functional conservation
  of takeout family members in a courtship model system. \emph{{PLoS} One}
  \textbf{2018}, \emph{13}, e0204615\relax
\mciteBstWouldAddEndPuncttrue
\mciteSetBstMidEndSepPunct{\mcitedefaultmidpunct}
{\mcitedefaultendpunct}{\mcitedefaultseppunct}\relax
\EndOfBibitem
\bibitem[Cherezov \latin{et~al.}(2007)Cherezov, Rosenbaum, Hanson, Rasmussen,
  Thian, Kobilka, Choi, Kuhn, Weis, Kobilka, and Stevens]{Cherezov:2007}
Cherezov,~V.; Rosenbaum,~D.; Hanson,~M.; Rasmussen,~S.; Thian,~F.; Kobilka,~T.;
  Choi,~H.; Kuhn,~P.; Weis,~W.; Kobilka,~B. \latin{et~al.}  High-resolution
  crystal structure of an engineered human beta2-adrenergic {G} protein-coupled
  receptor. \emph{Science} \textbf{2007}, \emph{318}, 1258--1265\relax
\mciteBstWouldAddEndPuncttrue
\mciteSetBstMidEndSepPunct{\mcitedefaultmidpunct}
{\mcitedefaultendpunct}{\mcitedefaultseppunct}\relax
\EndOfBibitem
\bibitem[Weiss and Schulz(1992)Weiss, and Schulz]{Weiss:1992}
Weiss,~M.; Schulz,~G. Structure of porin refined at 1.8 \AA\ resolution.
  \emph{J.\ Mol.\ Biol.} \textbf{1992}, 493--509\relax
\mciteBstWouldAddEndPuncttrue
\mciteSetBstMidEndSepPunct{\mcitedefaultmidpunct}
{\mcitedefaultendpunct}{\mcitedefaultseppunct}\relax
\EndOfBibitem
\bibitem[Bai(1992)]{Bai:1992}
Bai,~T. Beta 2 adrenergic receptors in asthma: a current perspective.
  \emph{Lung} \textbf{1992}, \emph{170}, 125--141\relax
\mciteBstWouldAddEndPuncttrue
\mciteSetBstMidEndSepPunct{\mcitedefaultmidpunct}
{\mcitedefaultendpunct}{\mcitedefaultseppunct}\relax
\EndOfBibitem
\bibitem[Graham \latin{et~al.}(2018)Graham, Leja, and Carlson]{Graham:2018}
Graham,~S.; Leja,~N.; Carlson,~H. {MixMD Probeview}: robust binding site
  prediction from cosolvent simulations. \emph{J.\ Chem.\ Inf.\ Model.}
  \textbf{2018}, \emph{58}, 1426--1433\relax
\mciteBstWouldAddEndPuncttrue
\mciteSetBstMidEndSepPunct{\mcitedefaultmidpunct}
{\mcitedefaultendpunct}{\mcitedefaultseppunct}\relax
\EndOfBibitem
\bibitem[Tan and Verma(2020)Tan, and Verma]{Tan:2020}
Tan,~Y.; Verma,~C. Straightforward incorporation of multiple ligand types into
  molecular dynamics simulations for efficient binding site detection and
  characterization. \emph{J.\ Chem.\ Theory Comput.} \textbf{2020}, \emph{16},
  6633--6644\relax
\mciteBstWouldAddEndPuncttrue
\mciteSetBstMidEndSepPunct{\mcitedefaultmidpunct}
{\mcitedefaultendpunct}{\mcitedefaultseppunct}\relax
\EndOfBibitem
\bibitem[Busch \latin{et~al.}(2011)Busch, Reijerse, Lubitz, Frankerberg-Dinkel,
  and Hofmann]{Busch:2011}
Busch,~A.; Reijerse,~E.; Lubitz,~W.; Frankerberg-Dinkel,~N.; Hofmann,~E.
  Structural and mechanistic insight into the ferredoxin-mediated two-electron
  reduction of bilins. \emph{Biochem.\ J.} \textbf{2011}, \emph{439},
  257--264\relax
\mciteBstWouldAddEndPuncttrue
\mciteSetBstMidEndSepPunct{\mcitedefaultmidpunct}
{\mcitedefaultendpunct}{\mcitedefaultseppunct}\relax
\EndOfBibitem
\bibitem[Dammeyer and Frankenberg-Dinkel(2008)Dammeyer, and
  Frankenberg-Dinkel]{Dammeyer:2008}
Dammeyer,~T.; Frankenberg-Dinkel,~N. Function and distribution of bilin
  biosynthesis enzymes in photosynthetic organisms. \emph{Photochem.\
  Photobiol.\ Sci.} \textbf{2008}, \emph{7}, 1121--1130\relax
\mciteBstWouldAddEndPuncttrue
\mciteSetBstMidEndSepPunct{\mcitedefaultmidpunct}
{\mcitedefaultendpunct}{\mcitedefaultseppunct}\relax
\EndOfBibitem
\bibitem[Frankenberg \latin{et~al.}(2001)Frankenberg, Mukougawa, Kohchi, and
  Lagarias]{Frankenberg:2001}
Frankenberg,~N.; Mukougawa,~K.; Kohchi,~T.; Lagarias,~J. Functional genomics
  analysis of the {HY2} family of ferredoxin-dependent bilin reductases from
  oxyhenic photosynthetic organisms. \emph{Plant Cell} \textbf{2001},
  \emph{13}, 965--978\relax
\mciteBstWouldAddEndPuncttrue
\mciteSetBstMidEndSepPunct{\mcitedefaultmidpunct}
{\mcitedefaultendpunct}{\mcitedefaultseppunct}\relax
\EndOfBibitem
\bibitem[Cheng \latin{et~al.}(2003)Cheng, Lowe, Sinclair, Nigg, and
  Johnson]{Cheng:2003}
Cheng,~K.-Y.; Lowe,~E.; Sinclair,~J.; Nigg,~E.; Johnson,~L. The crystal
  structure of the human polo-like kinase-1 polo box domain and its
  phospho-peptide complex. \emph{{EMBO} J.} \textbf{2003}, \emph{22},
  5757--5768\relax
\mciteBstWouldAddEndPuncttrue
\mciteSetBstMidEndSepPunct{\mcitedefaultmidpunct}
{\mcitedefaultendpunct}{\mcitedefaultseppunct}\relax
\EndOfBibitem
\bibitem[Liu \latin{et~al.}(2011)Liu, Park, Qian, Lim, Graeber, Berg, Yaffe,
  Lee, and {Burke Jr}]{Liu:2011}
Liu,~F.; Park,~J.; Qian,~W.-J.; Lim,~D.; Graeber,~M.; Berg,~T.; Yaffe,~M.;
  Lee,~K.; {Burke Jr},~T. Serendipitous alkylation of a {Plk1} ligand uncovers
  a new binding channel. \emph{Nat.\ Chem.\ Biol.} \textbf{2011}, \emph{7},
  595--601\relax
\mciteBstWouldAddEndPuncttrue
\mciteSetBstMidEndSepPunct{\mcitedefaultmidpunct}
{\mcitedefaultendpunct}{\mcitedefaultseppunct}\relax
\EndOfBibitem
\bibitem[\'{S}led\'{z} \latin{et~al.}(2011)\'{S}led\'{z}, Stubbs, Lang, Yang,
  MacKenzie, Venkitaraman, Hyv\"{o}nen, and Abell]{Sledz:2011}
\'{S}led\'{z},~P.; Stubbs,~C.; Lang,~S.; Yang,~Y.-Q.; MacKenzie,~G.;
  Venkitaraman,~A.; Hyv\"{o}nen,~M.; Abell,~C. From crystal packing to
  molecular recognition: prediction and discovery of a binding site on the
  surface of polo-like kinase 1. \emph{Angew.\ Chem.\ Int.\ Ed.} \textbf{2011},
  \emph{50}, 4003--4006\relax
\mciteBstWouldAddEndPuncttrue
\mciteSetBstMidEndSepPunct{\mcitedefaultmidpunct}
{\mcitedefaultendpunct}{\mcitedefaultseppunct}\relax
\EndOfBibitem
\bibitem[Tirion(1996)]{Tirion:1996}
Tirion,~M. Large amplitude elastic motions in proteins from a single parameter,
  atomic analysis. \emph{Phys.\ Rev.\ Lett.} \textbf{1996}, \emph{77},
  1905--1908\relax
\mciteBstWouldAddEndPuncttrue
\mciteSetBstMidEndSepPunct{\mcitedefaultmidpunct}
{\mcitedefaultendpunct}{\mcitedefaultseppunct}\relax
\EndOfBibitem
\bibitem[Sanejouand(2013)]{Sanejouand:2013}
Sanejouand,~Y. Elastic network models: theoretical and empirical foundations.
  \emph{Methods Mol.\ Biol.} \textbf{2013}, \emph{914}, 601--616\relax
\mciteBstWouldAddEndPuncttrue
\mciteSetBstMidEndSepPunct{\mcitedefaultmidpunct}
{\mcitedefaultendpunct}{\mcitedefaultseppunct}\relax
\EndOfBibitem
\bibitem[Sinitskiy and Voth(2013)Sinitskiy, and Voth]{Voth:2013}
Sinitskiy,~A.; Voth,~G. Coarse-graining of proteins based on elastic network
  models. \emph{Chem.\ Phys.} \textbf{2013}, \emph{422}\relax
\mciteBstWouldAddEndPuncttrue
\mciteSetBstMidEndSepPunct{\mcitedefaultmidpunct}
{\mcitedefaultendpunct}{\mcitedefaultseppunct}\relax
\EndOfBibitem
\end{mcitethebibliography}

\providecommand{\latin}[1]{#1}
\makeatletter
\providecommand{\doi}
  {\begingroup\let\do\@makeother\dospecials
  \catcode`\{=1 \catcode`\}=2 \doi@aux}
\providecommand{\doi@aux}[1]{\endgroup\texttt{#1}}
\makeatother
\providecommand*\mcitethebibliography{\thebibliography}
\csname @ifundefined\endcsname{endmcitethebibliography}
  {\let\endmcitethebibliography\endthebibliography}{}

\end{document}